\documentclass[aps,pra,twocolumn,groupedaddress,showpacs]{revtex4}

\usepackage{graphicx}
\usepackage{graphicx}
\usepackage{amssymb}
\usepackage{amsmath}
\usepackage{color}
\newcommand{\be}{\begin{equation}}
\newcommand{\ee}{\end{equation}}
 
 \definecolor{BrickRed}{cmyk}{0,0.89,0.94,0.28}
\definecolor{MidnightBlue}{cmyk}{0.98,0.13,0,0.43}
\definecolor{DarkGreen}{rgb}{0,0.7,0.1}

\begin{document}

\title{Casimir-Polder  shift of ground-state  hyperfine Zeeman  sub-levels of hydrogen isotopes in a micron-sized metallic cavity at finite temperature}


\author{Davide Iacobacci$^{1,2}$, Giuseppe Bimonte$^{1,2}$, and Thorsten Emig$^{3}$}

\affiliation{${}^{1}$Dipartimento di Fisica E. Pancini, Universit\`{a} di
Napoli Federico II, Complesso Universitario
di Monte S. Angelo,  Via Cintia, I-80126 Napoli, Italy}
\affiliation{${}^{2}$INFN Sezione di Napoli, I-80126 Napoli, Italy}
\affiliation{${ }^{3}$ Laboratoire de Physique
Th\'eorique et Mod\`eles Statistiques, CNRS UMR 8626,
Universit\'e Paris-Saclay, 91405 Orsay cedex, France}

\email{davide.iacobacci@unina.it, giuseppe.bimonte@na.infn.it,
thorsten.emig@universite-paris-saclay.fr}

\begin{abstract}

The frequencies of  transitions between  hyperfine levels of ground-state atoms  can be measured with exquisite precision using magnetic-resonance techniques. This makes hyperfine transitions ideal probes of QED effects originating from the interaction of atoms with the quantized electromagnetic field. One of the most remarkable effects predicted by QED is the Casimir-Polder shift  experienced by the  energy levels of  atoms placed near one or more dielectric objects.  Here we  compute the Casimir-Polder shift and the width of hyperfine transitions between ground-state Zeeman sub-levels of an hydrogen atom placed in a micron-sized metallic cavity,  over a  range of temperatures extending from cryogenic temperatures to room temperature.  Results are presented also for deuterium and tritium.  We predict shifts of the hyperfine transitions frequencies  of  a few tens of Hz that might be measurable with present-day magnetic resonance apparatus.

\end{abstract}


\maketitle

\section{Introduction}
\label{sec:intro}

A fundamental problem in atomic physics is the interaction of atoms with radiation fields. The atom's coupling with quantum fluctuations of the electromagnetic (em) field causes the spontaneous decay of excited states, and leads to the famous Lamb shift.  Studies carried out over the past seventy years or so revealed that  atomic level positions  are corrected by two further effects, that were not considered in the original Lamb shift.  On one hand, it was realized that  at finite temperature, besides virtual photons that are responsible of the Lamb shift, real photons give rise to a  temperature-dependent correction  of  atomic levels \cite{auluck,barton,knight,farley}.   
A different interesting situation is that of an atom in a confined geometry. Since the presence of  material boundaries   modifies  the spectrum of the modes of the em field, the lifetimes and energies of the atom's excited states are shifted \cite{wylie,kleppner}. Even though the ultimate explanation  of the latter phenomena is basically the same as the Lamb shift, i.e. coupling of the atom  with the modes of the em field, the term "Lamb shift" usually refers to  the shift of energy levels experienced by an atom in free space (at zero temperature). The additional correction to the energy levels resulting from the presence of one or more material surfaces (possibly at finite temperature)  is referred to as Casimir-Polder (CP) shift (for a recent review of the CP interaction, see the book \cite{buh} and references therein).

The observation of shifts of atomic energy levels induced by thermal radiation and/or by the CP interaction is difficult, because  the theoretically predicted magnitude of the shifts is extremely small. Over the past forty years, however,  several experiments  succeeded in observing these tiny effects, using high-precision laser spectroscopy techniques.  In \cite{hollberg} the shift of Rydberg energy levels of Rb atoms induced by black-body radiation was  measured in free space. The reported fractional shifts of $\sim 2 \times 10^{-12}$ were found to be consistent with theoretical expectations. A more recent experiment \cite{marrocco} observed  the shift of Rydberg energy levels  of Rb atoms confined between two parallel metal plates.   In this experiment the plate distance, of the order of a mm,  was tuned over the wavelength of the relevant Rydberg transitions. At resonance, level shifts of the order of 100 Hz were observed, for a range of temperatures extending from room temperature down to 4 K, again in good agreement with theory. A spectroscopic observation of the  CP shift of the levels of Rydberg  Na atoms in a non-resonant micron-sized cavity was  reported in \cite{haroche}. In the regime probed in this experiment, the CP interaction was observed in the non-retarded  limit, where temperature effects are negligible.  

In this paper, we  compute the thermal  CP shift of  the transition frequencies among ground-state hyperfine Zeeman sub-levels of  a hydrogen atom,  placed in a micron-sized metallic cavity at finite temperature. Besides hydrogen, we shall also consider its isotopes deuterium (D) and tritium (T). The present work is a follow-up of the companion paper \cite{bimonte2019}, where we investigated the influence of the CP interaction on the transition rates between hyperfine ground-state sub-levels of D atoms, passing between two closely spaced Au mirrors  at room temperature.  In that paper,  it was shown that the space between the mirrors   is filled with a broad non-Planckian spectrum of em noise, mainly consisting of thermal fluctuations of the magnetic field.  The energy density of these em fields is enormous:  in \cite{bimonte2019} it was estimated that at the center of a 2 $\mu$m Au cavity at room temperature, this density is about 700 times larger than the    energy density of a black-body at the same temperature. The existence of such a strong fluctuating magnetic field leads one to expect that the properties of the hyperfine sub-levels of an atom placed inside a cavity might be strongly affected, via coupling with the atom's magnetic moment. This expectation was indeed confirmed by the computations presented in \cite{bimonte2019}, where it was found that the transition rates 
between hyperfine sub-levels of D atoms placed at the center of a 2 $\mu$m Au cavity at room temperature   increase by approximately 15 orders of magnitude compared to the corresponding rates in a large black-body at the same temperature.  

In \cite{bimonte2019} we investigated the {\it absorptive} component of the CP atom-wall interaction,  which determines the transition rates among the hyperfine atomic levels. In the present work we  consider the {\it dispersive} component of the interaction to compute the frequency shift and width of the hyperfine transitions  of H atoms.
A preliminary investigation of this problem  was  sketched in \cite{carsten2}, where the shift of the energy of the $\rm 5 s$    $|{\rm F,m_F} \rangle=|{ 1,-1} \rangle$ ground-state hyperfine level of a Rb atom placed near a single metallic plate was estimated on the basis of a simplified two-level model of the  atom. In this paper the problem is studied  on the basis of a realistic model of the atom, which takes into account all its excited states.  

The paper is organized as follows. In Sec.~II we  review the perturbative theory of Wylie and Sipe \cite{wylie}  which can be used to compute the CP interaction of an atom with a material wall at finite temperature. In Sec.~III we  review some basic facts on the hyperfine structure of atomic spectra. These notions  are well-known  to atomic physicists. They are reviewed here for the benefit of readers working in the field of Casimir physics,  who may not be fully familiar with the properties of hyperfine states of atoms.  In Sec.~IV we compute the shift of hyperfine transitions induced by black-body radiation in free-space, while in Sec.~V we consider the case of an atom placed inside a cavity at finite temperature. In Sec.~VI we present our conclusions. Finally, in the Appendices we derive the  formulae for the matrix elements of the magnetic and electric dipole operators that are needed in the computation of the shift of hyperfine transitions, and we provide the explicit expression of the Green function of the electromagnetic field inside a dielectric cavity. Gaussian electromagnetic units are used throughout.

\section{Theory}

In this Section, we shall briefly review the basic  theory that allows to compute the shifts and the widths of the spectral lines of an atom placed at the point ${\bf r}_0$ near one or more material surfaces. The surfaces  are assumed to be in thermal equilibrium with the environment at temperature $T$.  

By working in the dipole approximation and using second-order perturbation theory, Wylie and Sipe \cite{wylie} showed that the shift of the energy of the state $| a \rangle$ of an atom exposed to thermal radiation can be decomposed as the sum of a dynamic Stark effect $\delta {\cal F}^{\rm (E)}_a$ induced by the atom's coupling to the thermal and quantum fluctuations of the electric  field,  plus a dynamic Zeeman  effect $\delta {\cal F}^{\rm (H)}_a$  induced by the atom's coupling to the thermal and quantum fluctuations of the magnetic  field,
\be
\delta {\cal F}_a=\delta {\cal F}^{\rm (E)}_a+\delta {\cal F}^{\rm (H)}_a\;,\label{splitSZ}
\ee  
where
\begin{eqnarray}
\!\!\delta {\cal F}_a^{(\rm E)} \!\!\!&\!=&\!\!\!-\frac{1}{\pi} \sum_b \!\!\!\!\!\!\!\!\int d^{a b}_{i} d^{ba}_{j} {\rm P} \!\!\int_{\!-\infty}^{\infty} \!\!\!\frac{d \omega\, {\rm Im} [{\cal E}_{i j}({\bf r}_0, {\bf r}_0; \omega )]}{(\omega+\omega_{ba}) (1-e^{-\hbar \omega/ k_B T})}, \\
 \!\!\delta {\cal F}_a^{(\rm H)} \!\!\!&\!=&\!\!\!-\frac{1}{\pi} \sum_b \!\!\!\!\!\!\!\!\int  \mu^{ab}_{i} \mu^{ba}_{j}  {\rm P}\!\!\int_{-\infty}^{\infty} \!\!\! \frac{ d \omega\;{\rm Im} 
[{\cal H}_{i j}({\bf r}_0, {\bf r}_0; \omega )]}{(\omega+\omega_{ba}) (1-e^{-\hbar \omega/ k_B T})}\, .\label{wylie}
\end{eqnarray}
In the above Equations, the symbol ${\rm P}$  denotes the principal value,  latin indices $i,j=1,2,3$ denote spatial directions, with repeated indices  summed over, and the symbol $\Sigma \,\!\!\!\!\!\!\int$   denotes   a sum over  discrete states, as well as in integral over continuum states. Moreover we define  the transition frequency
\be
\omega_{ba}=(E_b-E_a)/\hbar\,,
\ee 
where $E_a$ is  the unperturbed energy of state $a$, $d_{i}^{ab}= \langle a | {\hat d}_{i}| b \rangle$ and $\mu_{i}^{a b}= \langle a | {\hat \mu}_{i}| b \rangle$ are, respectively, the matrix elements of the atom's electric and magnetic dipole moment operators, and ${\cal E}_{i j}({\bf r}, {\bf r}'; \omega )$ and  ${\cal H}_{i j}({\bf r}, {\bf r}'; \omega )$ denote, respectively, the Fourier transformed \footnote{The Fourier transform $f(\omega)$ of a function $g(t)$ is defined such that $f(\omega)=\int_{-\infty}^{\infty} dt \;g(t)$} classical Green tensors of the electric and magnetic fields.   

In investigations of the CP interaction of atoms with material surfaces the contribution of the dynamic Zeeman effect $\delta {\cal F}^{\rm (H)}_a$ is usually negligible, and the entire interaction originates from the dynamic Stark effect. Contrary, in the problem studied in this paper, i.e. the CP shift of the transition frequencies among hyperfine Zeeman sub-levels of ground state atoms, the opposite true.  We shall find that the dominant role is indeed played by the dynamic Zeeman effect. 

As stated already, Eq.~(\ref{wylie}) is derived on the basis of second-order perturbation theory. An elegant non-perturbative computation of the energy shifts, based on the solution of the coupled dynamics of the  {\it macroscopic} quantized em field and the atomic system, is offered in \cite{buhman} (see also the book \cite{buh}). We note that the  non-perturbative result of  \cite{buhman} correctly reproduces Eq.~(\ref{wylie}) in the weak-coupling limit.  

In their paper, Wylie and Sipe actually used Eq.~(\ref{wylie}) for $T=0$, and they refer to $\delta {\cal F}_a$ as  the atom's `energy' shift, which is indeed correct at zero temperature. The careful analysis carried out by Barton \cite{barton1987}  shows that   for finite temperature $T$, the quantity $\delta {\cal F}_a$  actually represents the shift of the Helmholtz free-energy of the {\it constrained} equilibrium thermodynamic  system formed by the atom together with the quantized em field. Here, the atom's state label $a$  plays the same role as a macroscopically controlled variable in ordinary thermodynamics. For brevity, in what follows we shall  keep referring to $\delta {\cal F}_a$ as the energy-shift of state $a$.  According to Ref.~\cite{barton1987} the shifts $\delta \nu_{aa'}$ of the atom's transition frequencies $\nu_{aa'}=(E_a-E_{a'})/h$ (we assume $E_a>E_{a'}$ so that $\nu_{aa'}$ is defined to be positive) are calculable as  the differences between the coupling-induced  shifts of the free energies of the corresponding states,
\be
\;\;\;\;\;\;\;\delta \nu_{aa'}=(\delta {\cal F}_a-\delta {\cal F}_{a'})/h\;,\;\;\;\;(E_a > E_{a'}).\label{deltanu}
\ee
In view of our sign convention for the frequencies $\nu_{aa'}$, a positive $\delta \nu_{aa'}>0$ indicates that the transition frequency is shifted towards larger frequencies, while a negative $\delta \nu_{aa'}<0$ indicates a shift towards lower frequencies. We see from Eq.~(\ref{splitSZ}) that similarly to the energy-shift, the frequency shift $\delta \nu_{aa'}$  can be also expressed as the sum of the dynamic Stark shift $\delta \nu^{(\rm E)}_{aa'}$  and the dynamic Zeeman shift  $\delta \nu^{(\rm H)}_{aa'}$,
\be
\delta \nu_{aa'}=\delta \nu^{(\rm E)}_{aa'}+\delta \nu^{(\rm H)}_{aa'}\;,\label{totshift}
\ee 
where
\be
\delta \nu^{(\rm E)}_{aa'}=(\delta {\cal F}_a^{(\rm E)}-\delta {\cal F}_{a'}^{(\rm E)})/h\;,\label{shiftE}
\ee
\be
\delta \nu^{(\rm H)}_{aa'}=(\delta {\cal F}_a^{(\rm H)}-\delta {\cal F}_{a'}^{(\rm H)})/h\;.\label{shiftH}
\ee
Using first-order perturbation theory, Wylie and Sipe obtained the following expression for the transition probabilities  $A_{b a }$  of the allowed dipole  transitions from  state  $a$ to state  $b$,
\begin{eqnarray}
A_{b a}&=&\frac{2/\hbar}{ 1-e^{-\hbar \omega_{ab}/k_B T}} \left\{
d_{i}^{a b} d_{j}^{b a} \; {\rm Im}[{\cal E}_{i j}({\bf r}_0,{\bf r}_0,\omega_{ab})] \right.\nonumber\\
&+& \left.\mu_{i}^{a b }\mu_{j}^{b a} \; {\rm Im}[{\cal H}_{i j}({\bf r}_0,{\bf r}_0,\omega_{a b})] \right\}\, .\label{rate}
\end{eqnarray}
By summing the transition rate over all final states,   one obtains  the total  depopulation rate $\Gamma_a$,
\be
\Gamma_a = \sum_{b \neq a}\;A_{b a}\, .
\ee  
As it is well known \cite{corney},   the half-width $\Delta \omega_{1/2}$ of the spectral line corresponding to transitions between states $a$ and $b$ is the sum of the depopulation rates of the states $a$ and $b$,
\be
 \Delta \omega_{1/2}= \Gamma_b+\Gamma_a\, .\label{width}
\ee
We note that by virtue of the reciprocity relations   satisfied by the Green tensors ${\cal E}_{i j}({\bf r}, {\bf r}'; \omega )={\cal E}_{j i}({\bf r}', {\bf r}; \omega )$ and  ${\cal H}_{ i j}({\bf r}, {\bf r}'; \omega )={\cal H}_{j i}({\bf r}', {\bf r}; \omega )$ \cite{buh},  the products of the dipole moments  in Eqs.~(\ref{wylie}) and (\ref{rate})  can be actually replaced by their symmetrized products,
\begin{eqnarray}
 d^{ab}_{i} d^{b a}_{j}|_{\rm sym}&\equiv& \frac{d^{a b}_{i} d^{ki}_{j}+ d^{ab}_{j} d^{ba}_{i}}{2}\;,\nonumber\\
 \mu^{ab}_{i} \mu^{ba}_{j}|_{\rm sym}&\equiv& \frac{\mu^{ab}_{i} \mu^{ba}_{j}+ \mu^{ab}_{j} \mu^{ba}_{i}}{2}\, .
\end{eqnarray}

Eqs.~(\ref{wylie}) and (\ref{rate}) provide the general framework for studying the  shifts and the radiative widths of the spectral lines of an atom which interacts with the radiation field surrounding one or more dielectric bodies, in thermal equilibrium with the environment. In the following subsection, we shall study first the simple situation of an atom exposed to black-body radiation in free-space, while in subsection B we shall consider the more complicated case of an atom placed inside a cavity. 

\subsection{An atom in free space}

 The  shifts and the radiative widths of the  spectral lines of an atom, exposed to black-body radiation  in free-space, can be computed by substituting  into Eqs~.(\ref{wylie}) and (\ref{rate}) the free-space Green tensors ${\cal E}^{(0)}_{ij}({\bf r}, {\bf r}'; \omega )$ and  ${\cal H}^{(0)}_{ij}({\bf r}, {\bf r}'; \omega )$. The expressions of the latter Green tensors are provided in Appendix B. It can be seen from Eq.~(\ref{Green0}) that  in the coincidence limit ${\bf r}={\bf r}'={\bf r}_0$ the real parts of the free-space Green tensors diverge, while their imaginary parts attain the finite limit
\be
{\rm Im}[{\bf \cal E}_{ij}^{(0)}({\bf r}_0,{\bf r}_0;\omega)]={\rm Im}[{\bf \cal H}_{ij}^{(0)}({\bf r}_0,{\bf r}_0;\omega)]=\frac{2 \omega^3}{3 c^3}\;\delta_{ij}\, .
\ee
When the latter formula is substituted back into Eq.~(\ref{wylie}) it is not hard to verify that the energy-shift
$\delta {\cal F}_a^{(0)}$ can be decomposed into the sum of two terms,
\be
\delta {\cal F}_a^{(0;{\rm bare})}=\delta {\cal F}_a^{(0;{\rm zp})}+\delta {\cal F}_a^{(0;{\rm th})}\, ,\label{splitdelta}
\ee
where we have defined the zero-point contribution
\be
\label{zp0}
\delta {\cal F}_a^{(0;{\rm zp})}=-\frac{2}{3 \pi c^3}  \sum_b \!\!\!\!\!\!\!\!\int \left( d^{ab}_{i} d^{ba}_{i} +\mu^{ab}_{i} \mu^{ba}_{i}\right) 
  {\rm P}\!\!\!\int_{0}^{\infty} \!\! \! \!\!\frac{d \omega \, \omega^3}{(\omega_{ba}+\omega) }\;.
\ee
and the thermal shift
\be
\delta {\cal F}_a^{(0;{\rm th})}(T)=\delta {\cal F}_a^{(0;{\rm th|E})}(T)+\delta {\cal F}_a^{(0;{\rm th|H})}(T)\;,
\ee
with
\begin{eqnarray}
\delta {\cal F}_a^{(0;{\rm th|E})}(T)&=&-\frac{2 \,\hbar}{3 \pi c^3}  {\rm P}\!\int_{0}^{\infty} \!\! d \omega \frac{  \omega^3 \;  \alpha^{(a)}_{ii}(\omega) }{(e^{\hbar \omega/k_B T}-1)} \;, \nonumber\\
\delta {\cal F}_a^{(0;{\rm th|H})}(T) &=&-\frac{2 \,\hbar}{3 \pi c^3}  {\rm P}\!\int_{0}^{\infty} \!\! d \omega \frac{\omega^3 \;  \beta^{(a)}_{ii}(\omega) }{(e^{\hbar \omega/k_B T}-1)} \;.
\end{eqnarray}
Here,  $\alpha^{(a)}_{ij}(\omega)$ and $\beta^{(a)}_{ij}(\omega)$ denote, respectively,   the electric and magnetic polarizabilities  of the atom in the state $a$ \cite{bonin},
\begin{eqnarray}
\alpha^{(a)}_{ij}(\omega)&=& \frac{1}{\hbar}  \sum_{b \neq a} \!\!\!\!\!\!\!\!\int \left( \frac{d_i^{ab} d_j^{ba}}{\omega_{ba}-\omega} + \frac{d_j^{ab} d_i^{ba}}{\omega_{ba} +\omega} \right) \;, \label{defelpol}\\
\beta^{(a)}_{ij}(\omega)&=&  \frac{1}{\hbar}  \sum_{b \neq a} \!\!\!\!\!\!\!\!\int \left( \frac{\mu_i^{ab} \mu_j^{ba}}{\omega_{ba}-\omega} + \frac{\mu_j^{ab} \mu_i^{ba}}{\omega_{ba} +\omega} \right) \;.\label{defmagpol}
\end{eqnarray}
We  note that the contribution $\delta {\cal F}_a^{(0;{\rm zp})}$ in Eq.~(\ref{zp0}) is expressed by  an UV-divergent integral over frequency,  which is independent of temperature. This allows to interpret $\delta {\cal F}_a^{(0;{\rm zp})}$ as representing the formally divergent shift of the atom's energy engendered by the atom's coupling to vacuum fluctuations of the em field, i.e. its Lamb shift. Clearly, the non-relativistic theory behind Eq.~(\ref{wylie}) cannot properly account for this effect, whose computation requires consideration  of the full relativistic quantum theory. One can thus neglect  $\delta {\cal F}_a^{(0;{\rm zp})}$,  and   just retain  the second term $\delta {\cal F}_a^{(0;{\rm th})}(T)$  in Eq.~(\ref{splitdelta}), which provides the temperature-dependent energy shift of an atom interacting with a thermal bath,
\be
\delta {\cal F}_a^{(0)}=\delta {\cal F}_a^{(0;{\rm th})}(T)\;.\label{shiftfree}
\ee    
An Equation equivalent to Eq.~(\ref{shiftfree}) was indeed derived in Ref.~\cite{farley}, where it was used to compute the dynamic Stark  shifts $\delta {\cal F}_a^{(0;{\rm th|E})}(T)$ of Rydberg states of hydrogen, helium and alkali-metal atoms induced by black-body radiation. Following this reference, we  recast $\delta {\cal F}_a^{(0;{\rm th|E})}(T)$ and $\delta {\cal F}_a^{(0;{\rm th|H})}(T)$ in the form
\begin{eqnarray}
\!\!\!\!\!\!\!\!\!&&\delta {\cal F}_a^{(0;{\rm th|E})}\!\!=\!\frac{2}{3 \pi c^3}\! \left(\frac{k_B T}{\hbar} \right)^3   
\!\! \sum_{b \neq a} \!\!\!\!\!\!\!\!\int \sum_i  |d_i^{ab} |^2 F\left(\!\frac{E_a-E_b}{k_B T}\!\right),\nonumber \\
\!\!\!\!\! \!\!\!\!&&\delta {\cal F}_a^{(0;{\rm th|H})}\!\!=\!\frac{2}{3 \pi c^3}\! \left(\frac{k_B T}{\hbar} \right)^{\!3}   \!\!
 \sum_{b \neq a} \!\!\!\!\!\!\!\!\int \sum_i  |\mu_i^{ab} |^2 F\left(\!\frac{E_a-E_b}{k_B T}\!\right) \, ,
\label{farleyfor}
\end{eqnarray}
where the function $F$ is given by
\be
F(y)={\rm P}\int_0^{\infty} d x \left(\frac{1}{y+x}+\frac{1}{y-x} \right) \frac{x^3}{e^x-1}\;.\label{farley}
\ee
\begin{figure}[ht]
\includegraphics [width=.9\columnwidth]{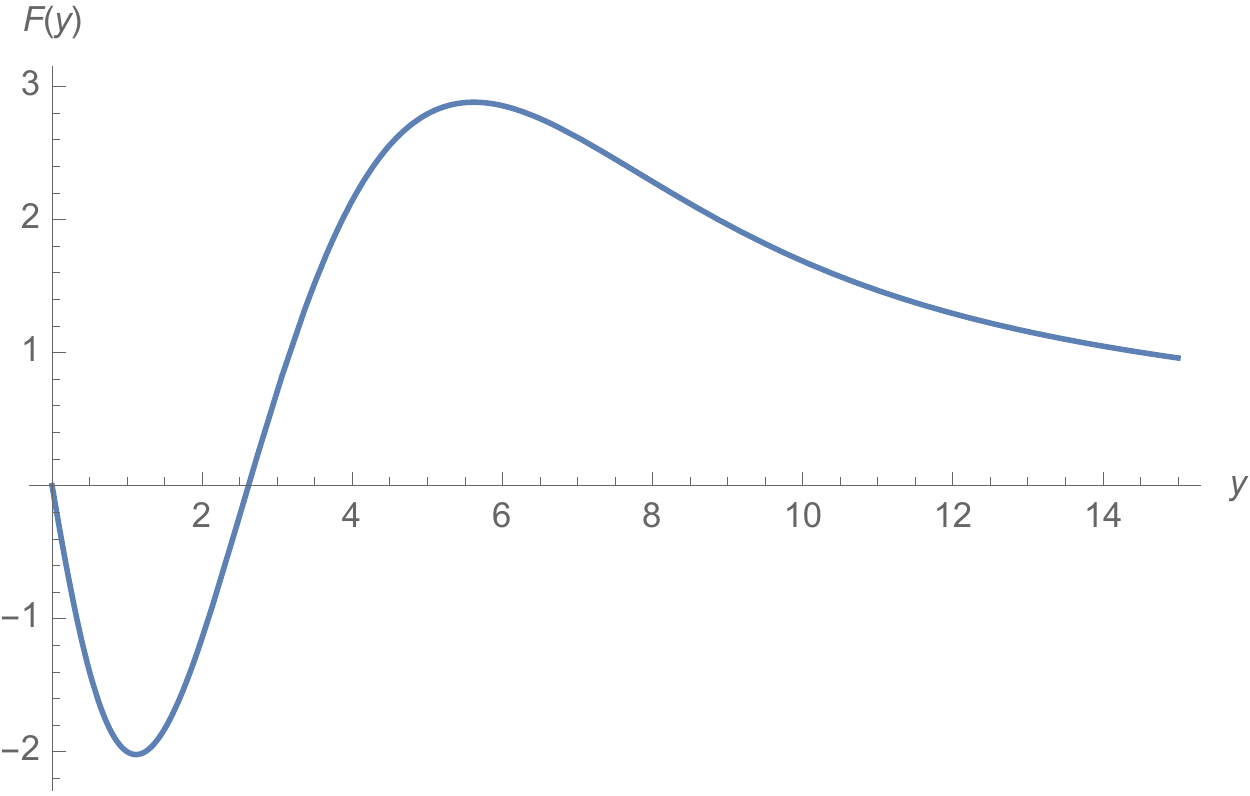}
\caption{\label{Fig1}  Graph of the function $F(y)$ defined in Eq.~(\ref{farley})}
\end{figure}
$F(y)$, which is an odd function of $y$, is displayed in Fig.\ref{Fig1}. In Ref.~\cite{farley} it is shown that $F(y)$ has the following asymptotic behavior for small and large  values of $|y|$,
\be
F(y) \simeq \left\{ \begin{array}{ll}
\!-\frac{\pi^2 y}{3} & |y|\ll1\\
\frac{2 \pi^4}{15 y}+\frac{16 \pi^6}{63 y^3} & |y|\gg1 \\
\end{array}\right. \, .\label{asF}
\ee
The frequency shift $\delta \nu_{aa'}^{(\rm 0;th)}$ of an atom exposed to black-body radiation can be now be written as
\be
\delta \nu_{aa'}^{(\rm 0;th)}= \delta \nu_{aa'}^{(\rm 0;th|E)}+\delta \nu_{aa'}^{(\rm 0;th|H)}\;,\label{totalshift0}
\ee
where
\begin{eqnarray}
 \delta \nu_{aa'}^{(\rm 0;th|E)}&=&(\delta {\cal F}_a^{(0;{\rm th|E})}-\delta {\cal F}_{a'}^{(0;{\rm th|E})})/h\;,\label{shiftE0}\\
 \delta \nu_{aa'}^{(\rm 0;th|H)}&=&(\delta {\cal F}_a^{(0;{\rm th|H})}-\delta {\cal F}_{a'}^{(0;{\rm th|H})})/h\;.\label{shiftH0}
 \end{eqnarray}
In Sec.~V the above equations will be used to compute the shift of 
hyperfine transitions in free-space.

\subsection{An atom near one or more bodies}

The shifts and  the radiative widths of the  spectral lines  of an atom in the presence of one or more bodies in thermal equilibrium with the environment can be computed using Eqs.~(\ref{wylie}) and (\ref{rate}), provided that the appropriate Green tensors are substituted into both equations.  In systems involving a collection of bodies, the Green tensors are naturally decomposed as the sum of the free-space Green tensors, and a scattering contribution which accounts for the influence of the bodies,
\begin{eqnarray}
{\cal E}_{ij}({\bf r},{\bf r'},\omega)&=&{\cal E}_{ij}^{(0)}({\bf r},{\bf r'},\omega)+{\cal E}^{(\rm sc)}_{ij}({\bf r},{\bf r'},\omega)\;,\nonumber \\
{\cal H}_{ij}({\bf r},{\bf r'},\omega)&=&{\cal H}_{ij}^{(0)}({\bf r},{\bf r'},\omega)+{\cal H}^{(\rm sc)}_{ij}({\bf r},{\bf r'},\omega)\;.\label{splitGr}
\end{eqnarray}
We note that, differently from the free-space contribution, the scattering parts of the Green tensors are non-singular in the coincidence limit ${\bf r}={\bf r}'={\bf r}_0$.
Unfortunately, they are difficult to compute for general shapes and dispositions of the bodies, and their expressions are known   only for a handful of special geometries.  In  the simple setting of two plane-parallel indefinite slabs the Green tensors  can be  written in terms of the Fresnel reflection coefficients of the slabs. Their explicit  expressions are provided in Appendix B.  The decomposition Eq.~(\ref{splitGr})  implies that the energy shift $\delta {\cal F}_a$  can be also represented as the sum of  two terms,
\be
\delta {\cal F}_a=\delta {\cal F}^{(0;{\rm bare})}_a+\delta {\cal F}^{(\rm sc)}_a \, ,
\ee   
where $\delta {\cal F}^{(0;{\rm bare})}_a$ is identical with the free-space shift in Eq.~(\ref{splitdelta}) and $\delta {\cal F}^{(\rm sc)}_a$ has an expression analogous to Eq.~(\ref{wylie}), apart from the replacement of the Green tensors with their scattering parts. After neglecting the divergent contribution from zero-point fluctuations  $\delta {\cal F}_a^{(0;{\rm zp})}$, one arrives at the following expression for the energy shift,
\be
\delta {\cal F}_a=\delta {\cal F}^{(0;{\rm th})}_a+\delta {\cal F}^{(\rm sc)}_a\, .\label{splitdeltaF}
\ee 
The scattering contribution $\delta {\cal F}^{(\rm sc)}_a$ can be further decomposed as the sum of the dynamic Stark shift and the dynamic Zeeman shift,
\be
\delta {\cal F}^{(\rm sc)}_a= \delta {\cal F}^{(\rm sc|E)}_a+\delta {\cal F}^{(\rm sc|H)}_a\, .
\ee
By taking advantage of the analyticity and symmetry properties of the Green tensors \cite{buh}, it is possible to show that 
$\delta {\cal F}^{(\rm sc|E)}_a$ and $\delta {\cal F}^{(\rm sc|H)}_a$  can be recast in the following form \cite{gorza},
\begin{eqnarray}
\delta {\cal F}^{(\rm sc|E)}_a&=&\delta {\cal F}^{(\rm  abs|E)}_a +\delta {\cal F}^{(\rm em|E)}_a+\delta {\cal F}^{(\rm non\, res|E)}_a \, , \label{deltaFsc0E} \\ 
\delta {\cal F}^{(\rm sc|H)}_a&=&\delta {\cal F}^{(\rm  abs|H)}_a +\delta {\cal F}^{(\rm em|H)}_a+\delta {\cal F}^{(\rm non\, res|H)}_a,
\label{deltaFsc0}
\end{eqnarray}
where 
\begin{widetext}
\begin{eqnarray}
&&\delta {\cal F}^{(\rm abs|E)}_a=
\sum_{b > a} \!\!\!\!\!\!\!\!\int d^{a b}_{i} d^{ba}_{j}  \frac{ {\rm Re} [{\cal E}^{(\rm sc)}_{i j}({\bf r}_0, {\bf r}_0; \omega_{ba} )] }{e^{\hbar \omega_{ba}/k_B T}-1} \theta(\omega_{ba})\;,\label{absE}\\
&&\delta {\cal F}^{(\rm em|E)}_a=-  \sum_{b<a} d^{a b}_{i} d^{ba}_{j}  {\rm Re} [{\cal E}^{(\rm sc)}_{i j}({\bf r}_0, {\bf r}_0; \omega_{ab} )]  \frac{e^{\hbar \omega_{ab}/k_B T}}{ e^{\hbar \omega_{ab}/k_B T}-1} \theta(\omega_{ab}) \,,\label{emE}\\
&&\delta {\cal F}^{(\rm non\,res|E)}_a= -k_B T \sum_{n=0}^{\infty}\;\!\!'  {\cal E}^{(\rm sc)}_{i j}({\bf r}_0, {\bf r}_0; {\rm i} \,\xi_n )\,\alpha^{(a)}_{ij}({\rm i} \,\xi_n)  \; , \label{nonresE} \\
&&\delta {\cal F}^{(\rm abs|H)}_a=
\sum_{b > a} \!\!\!\!\!\!\!\!\int  
\mu^{a b}_{i} \mu^{ba}_{j}    \frac{{\rm Re} [{\cal H}^{(\rm sc)}_{i j}({\bf r}_0, {\bf r}_0; \omega_{ba} )] }{e^{\hbar \omega_{ba}/k_B T}-1} \theta(\omega_{ba})\;, \label{absH}\\ 
&&\delta {\cal F}^{(\rm em|H)}_a=-  \sum_{b<a} 
\mu^{a b}_{i} \mu^{ba}_{j}  {\rm Re} [{\cal H}^{(\rm sc)}_{i j}({\bf r}_0, {\bf r}_0; \omega_{ab} )] \frac{e^{\hbar \omega_{ab}/k_B T}}{ e^{\hbar \omega_{ab}/k_B T}-1} \theta(\omega_{ab}) \,,\label{emH}\\
&&\delta {\cal F}^{(\rm non\,res|H)}_a= -k_B T \sum_{n=0}^{\infty}\;\!\!' {\cal H}^{(\rm sc)}_{i j}({\bf r}_0, {\bf r}_0; {\rm i} \,\xi_n )\,\beta^{(a)}_{ij}({\rm i} \,\xi_n) \, . \label{deltaFsc1}
\end{eqnarray}
\end{widetext}

In the above Equations, $\theta(x)$ is the Heaviside function ($\theta(x)=1$ for $x \ge 0$, and $\theta(x)=0$ for $x<0$), $\xi_n=2 \pi n k_B T/\hbar$, with $n=0,1,\cdots$, are the Matsubara frequencies, and the primed sums in Eqs.~(\ref{nonresE}) and (\ref{deltaFsc1}) indicate that the $n=0$ term is taken with a weight of one-half. The three contributions on the r.h.s. of Eqs.~(\ref{deltaFsc0E})  and (\ref{deltaFsc0}) have different physical meanings: the first terms, $\delta {\cal F}^{(\rm abs|E)}_a$ and $\delta {\cal F}^{(\rm abs|H)}_a$ respectively,  correspond to virtual   absorption processes by the atom, and thus they only involve virtual intermediate states $b$ such that $E_b>E_a$. We note that  $\delta {\cal F}^{(\rm abs|E)}_a$ and $\delta {\cal F}^{(\rm abs|H)}_a$ vanish for $T=0$.    The second terms, $\delta {\cal F}^{(\rm em|E)}_a$ and $\delta {\cal F}^{(\rm em|H)}_a$, correspond to virtual    stimulated and spontaneous emission processes by the atom, and thus they only involve virtual intermediate states $b$ such that $E_b<E_a$. Finally, the third contributions, $\delta {\cal F}^{(\rm non\,res|E)}_a$ and  $\delta {\cal F}^{(\rm non\,res|H)}_a$, are associated with non-resonant quantum and thermal fluctuations of the atomic dipoles.  The resonant contributions $\delta {\cal F}^{(\rm abs)}_a$ and $\delta {\cal F}^{(\rm em)}_a$  represent truly  non-equilibrium contributions which exist only for atoms in non-thermalized states. In fact, it is  possible to verify that the resonant terms $\delta {\cal F}^{(\rm abs)}_a$ and $\delta {\cal F}^{(\rm em)}_a$ both cancel out when the atom's state is fully thermalized. This can be seen from the relation
$$
\delta {\cal F}^{(\rm eq)}_{\rm CP}\equiv\!\!  \frac{1}{Z}\sum_{ a} \!\!\!\!\!\!\!\!\int e^{-E_a/k_B T}\delta {\cal F}^{(\rm sc)}_a \! =\!  \frac{1}{Z}\sum_{a} \!\!\!\!\!\!\!\!\int e^{-E_a/k_B T}\delta {\cal F}^{(\rm non\,res)}_a\nonumber
$$
\be
= -k_B T \sum_{n=0}^{\infty}\;\!\!'   \left[ {\cal E}^{(\rm sc)}_{i j}( {\rm i} \,\xi_n )\,\alpha_{ij}({\rm i} \,\xi_n)  + {\cal H}^{(\rm sc)}_{i j}( {\rm i} \,\xi_n )\,\beta_{ij}({\rm i} \,\xi_n) \right],\;\;\label{shifteq} 
\ee
where 
\be
Z= \sum_{a} \!\!\!\!\!\!\!\!\int \exp(-E_a/k_B T)
\ee 
is the partition function (for brevity we omitted indicating the dependence of the Green tensors on the atom's position ${\bf r}_0$), and 
the polarizabilities of the thermalized atom are given by
\begin{eqnarray}
\alpha_{ij}({\rm i} \,\xi_n)&=&  \frac{1}{Z} \sum_{a} \!\!\!\!\!\!\!\!\int e^{-E_a/k_B T}\;\alpha^{(a)}_{ij}({\rm i} \,\xi_n)\;,\\
\beta_{ij}({\rm i} \,\xi_n)&=&  \frac{1}{Z} \sum_{ a} \!\!\!\!\!\!\!\!\int e^{-E_a/k_B T}\;\beta^{(a)}_{ij}({\rm i} \,\xi_n)
\end{eqnarray}
 Eq.~(\ref{shifteq}) shows that when the atom is in thermal equilibrium, the (scattering contribution to the)  atom's  free-energy shift $\delta {\cal F}^{(\rm eq)}_{\rm CP}$  coincides with the CP energy predicted by Lifshitz theory \cite{lifs}.   Using the expression of the Green tensors inside a cavity provided in Appendix B, it is possible to verify that Eq.~(\ref{shifteq}) indeed reproduces the formula  for the interaction 
 of a thermalized atom with a dielectric wall when the atom has a permanent magnetic moment. This result  was derived  in Ref.~\cite{bimonte2009} by taking the dilute limit of the Lifshitz formula. The cancellation of all resonant contributions for atoms in thermal equilibrium, and the importance of a proper interpretation of the atomic polarizability have been emphasized in \cite{buhman}. This becomes important when one studies non-equilibrium problems, like the CP interaction of an atom prepared in an energy eigenstate.

\section{Hyperfine structure}

In this Section, we briefly review the hyperfine structure of atomic spectra \cite{corney,gottfried}.  We  consider atoms with one optical electron, i.e., an hydrogenic atom, or an atom with just one electron outside closed shells, like an alkali atom.  The Hamiltonian ${\hat H}$  of such an atom  is given by
\be
{\hat H}={\hat H}_0+{\hat H}_{\rm HFS}\;,
\ee
where ${\hat H}_0$ describes the central electrostatic field of the atom and the spin-orbit interaction, while ${\hat H}_{\rm HFS}$ is the hyperfine  Hamiltonian. It is usually possible to treat ${\hat H}_{\rm HFS}$ as a small perturbation to ${\hat H}_0$, and then  the action  of  ${\hat H}_{\rm HFS}$ can be restricted to the subspace spanned by the electronic states $(n{\rm L} {\rm J} )$   having fixed principal quantum number $n$,  orbital quantum number ${\rm L}$, and  total electron angular momentum ${\rm J}$.  Within each $(n{\rm L J})$ sub-shell ${\hat H}_{\rm HFS}$  is of the form \cite{corney}
\begin{widetext}
\be
{\hat H}_{\rm HFS}=A_{\rm J} {\hat {\bf I}}\cdot  {\hat {\bf J}}+\frac{B_{\rm J}}{2 {\rm I}  (2 {\rm I} -1) {\rm  {\rm J} } (2  {\rm J} -1)} \left[3 \left({\hat {\bf I}}\cdot  {\hat {\bf J}}\right)^2+ \frac{3}{2}\, {\hat {\bf I}}\cdot  {\hat {\bf J}}-{\rm I} ({\rm I} +1) {\rm J(J+1)} \right] - \hat{\boldsymbol{\mu}} \cdot {\bf B}\;,\label{HHFS}
\ee
\end{widetext}
where  ${\bf B}$ is the external magnetic field, and
\be
\hat{\boldsymbol{\mu}}=-g_{\rm J} \mu_{\rm B} {\hat {\bf J}} + g_{\rm I} \mu_{\rm n} {\hat {\bf I}}\;.
\ee
is the atom's magnetic moment. Here, $ {\hat {\bf I}}$ and    $ {\hat {\bf J}}= {\hat {\bf L}}+ {\hat {\bf S}}$ are, respectively,  the  nuclear spin and the total angular momentum of the electron (both expressed in units of $\hbar$),  $\mu_{\rm B}=e \hbar/2 m c$ and $\mu_{\rm n}=e \hbar/2 M c$ are, respectively, the Bohr and nuclear magnetons, $m$ and $M$ are, respectively, the electron and the proton masses, $g_{\rm J} = [3 {\rm J(J+1)}+3/4-{\rm L} ({\rm L}  +1)]/2 {\rm J(J+1)}$ is the electron Land{\'e} g-factor (we use for the electron gyromagnetic factor the approximate value $g=2$),  $g_{\rm I}$  is the nuclear g-factor, $A_{\rm J}$ is the magnetic hyperfine structure constant,  and $B_{\rm J}$ is the electronic quadrupole interaction constant. For hydrogen the constant $A_J$  has the value \cite{gottfried},
\be
A_{\rm J}= \,w_0\,g_{\rm I}\, \frac{m}{M} \,\frac{ \alpha_{\rm e}^2}{n^3}\,  \frac{\rm 1}{\rm J(J+1)( {\rm L}+1/2)} \;,
\ee 
where $w_0$ is the Bohr energy.
The constant $B_{\rm J}$ is identically zero either if ${\rm I}=0$ or $1/2$, or if $ {\rm J} =0,1/2$ \cite{corney}. The latter condition implies, in particular, that $B_{\rm J}$ is zero in the ground states of the one-electron atoms that we consider.    

If the external magnetic field is zero,  the total angular momentum ${\rm F}$ and its projection ${\rm M_F}$ in any direction are good quantum numbers, and then the atom's states can be labelled by $|a \rangle = |n {\rm L} {\rm S}{\rm J}{\rm I};{\rm F} {\rm M_F} \rangle$, where $n_a={\rm L}+n_r+1$ ($n_r=0,1,2\dots$) is the principal quantum number, and ${\rm S}=1/2$ is the electron spin. For brevity, we shall omit  from now on the quantum numbers ${\rm S}$ and ${\rm I}$, and hence the states of the atom in zero external magnetic field shall be denoted simply as $|a \rangle = |n {\rm L} {\rm J};{\rm F} {\rm M_F} \rangle$. According to Eq.~(\ref{HHFS}) the energy $E_{\rm F}$ of this state is
\be
E_{\rm F}=E_{\rm J} + \frac{1}{2} A_{\rm J} {\rm K}+B_{\rm J} \frac{3 {\rm K(K+1)}-4 {\rm I(I+1)J(J+1)}}{8\, {\rm I}\,  (2 {\rm I}  -1)\,{\rm J\,(2 J-1)}}\;,\label{EFnoB}
\ee 
where $E_{\rm J}$ is the energy of the fine-structure multiplet level with electronic angular momentum ${\rm J}$, and ${\rm K}$ is
\be
{\rm K=F(F+1)-{\rm I} ({\rm I} +1)-J(J+1)}\;.
\ee
In our computations, we  approximate  $E_{\rm J}$ to  order $\alpha_{\rm e}^2$ \cite{messiah},
\be
E_{\rm J}= -w_0 \left[\frac{1}{n^2}+ \frac{\alpha_{\rm e}^2}{n^3} \left(\frac{1}{\rm J+1/2} -\frac{3}{4 n}\right) \right]\;,\label{finestr}
\ee 
where $\alpha_{\rm e}^2$ is the fine structure constant.
According to Eq.~(\ref{EFnoB}) the hyperfine interaction splits the fine-structure levels into  hyperfine-structure multiplets, consisting of a number of levels equals to $2{\rm I}+1$ if ${\rm J} \ge {\rm I} $ and $2{\rm J}+1$ if ${\rm I \ge J}$. 
For $B_{\rm J} \ll A_{\rm J}$ (and in particular in the ground state where $B_{\rm J}=0$) the adjacent hyperfine sub-levels are spaced  by the hyperfine-structure interval $\Delta E$
\be
\Delta E=E_{\rm F}-E_{{\rm F}-1}=A_{\rm J} F\;.
\ee 
When the external magnetic field ${\bf B}$ is different from zero, the degeneracy of the energy with respect to ${\rm M_F}$ is lifted, and the properties of the   Zeeman sub-levels  depend on the strength of ${\bf B}$. For weak fields such that $g_{\rm J} \mu_{\rm B} B \ll A_{\rm J}$ (this condition is typically satisfied if $B$ is less than $10^{-3}$ T),    the nuclear spin ${\rm I}$ and the electronic angular momentum ${\rm J}$ remain strongly coupled, and then the atom's eigenstates can be still labelled as $|a \rangle \equiv |n{\rm L}{\rm J}; {\rm F}{\rm M}_{\rm F} \rangle$, where ${\rm M_F}$ is the projection of the total angular momentum ${\bf F}$ in the direction of ${\bf B}$. The Zeeman effect adds to Eq. (\ref{EFnoB})    an energy $E_{\rm F M_F}$
\be
E_{\rm F M_F}= g_{\rm F} \mu_{\rm B} B\, {\rm M_F}\;,\label{enZee}
\ee 
where $g_{\rm F}$ is the effective g-value \cite{corney}. 
If the magnetic field $B$ does not satisfy the weak-field condition  $g_{\rm J} \mu_{\rm B} B \ll A_{\rm J}$, the energy levels must be determined by resolving the secular equation for ${\hat H}_{\rm HFS}$.  
While no  general formula can be written for arbitrary values of ${\rm I}$ and ${\rm J}$, the special case when either ${\rm I}$ or ${\rm J}$ do not exceed 1/2 can be solved quite easily. This case is of course important, because it applies to the $\;\!\!^2 S_{1/2}$ ground states of hydrogen and all alkali atoms. If the Zeeman sub-levels  are   labelled by  the quantum numbers $({\rm F, M_F})$ of the corresponding weak-field states,   the  energy is given by the Breit-Rabi formula
\begin{eqnarray}
E_{\rm F M_F}&=&-\frac{ h \nu_{\rm HFS} }{2(2{\rm I}+1)}-g_{\rm I} \mu_{\rm n} B {\rm M_F}  \nonumber \\ 
&\pm& \frac{h \nu_{\rm HFS}}{2} \left[1+\frac{4\, {\rm M_F} \,x}{2I+1} + x^2 \right]^{1/2}\;,\label{BR}
\end{eqnarray}
where 
\be 
h \nu_{\rm HFS}=A_{\rm J} ({\rm I}+1/2)
\ee 
is the energy separation between the ground-state sub-levels ${\rm F=I} \pm 1/2$ in zero magnetic field, and the dimensionless parameter $x$ is defined as
\be
x=\frac{(g_{\rm J} + g_{\rm I} m/M) \mu_{\rm B} B}{h \nu_{\rm HFS}}\;.
\ee
In Eq.~(\ref{BR}) the plus sign applies to states originating from the zero-field level ${\rm F=I}+1/2$, while the minus sign applies to states originating from the level ${\rm F=I}-1/2$.
\begin{table}[h]
\begin{tabular}{cccc} \hline 
Isotope & I & $\nu_{\rm HFS}$(Hz) & $\;\;\;\;g_{\rm I}$  \\ \hline \hline
$\!\!^1{\rm H}$ &\;1/2\;\;  &  1420405751.768 $\pm$ 0.001      &5.585486(0)    \\  \hline
 $\!\!^2{\rm H}$ & \;1  &   327384352.522 $\pm$ 0.002      &\;\;0.8574073(2)  \\  \hline
 $\!\!^3{\rm H}$ &1/2 & 1516701470.773 $\pm$ 0.008  &\;5.95768(2)   \\ 
\hline 
\end{tabular}
\caption{Values of the nuclear spin ${\rm I}$, energy separation $\nu_{\rm HFS}$ (in Hz) \cite{kars} and nuclear g-factor $g_{\rm I}$ for the three isotopes of hydrogen \cite{ramsey}.}
\label{tab.1}
\end{table}

In the next Sections we  shall compute the frequency shift $\delta \nu_{aa'}$ of the transitions between ground-state hyperfine sub-levels of  atoms exposed to thermal radiation, both in free-space and inside a metallic cavity.

\section{Shift and width of hyperfine transitions induced by black-body radiation}

In this Section we compute the shift $\delta \nu_{aa'}^{(0;{\rm th})}$  of the transition frequencies between  ground state hyperfine sub-levels of hydrogen and of its isotopes, that arise when the atom is bathed by black-body radiation in free space. 

According to Eq.~(\ref{totalshift0})  the shift $\delta \nu_{aa'}^{(0;{\rm th})}$ is the sum of two contributions,  the dynamic Stark shift $\delta \nu_{aa'}^{(0;{\rm th|E})}$ and the dynamic Zeeman shift  $\delta \nu_{aa'}^{(0;{\rm th|H})}$.
We consider first the  dynamic Zeeman shift.  We see from Eqs.~(\ref{farleyfor}) and (\ref{shiftH0}) that $\delta \nu_{aa'}^{(0;{\rm th|H})}$ involves  transitions to intermediate states $b$ that are coupled by the atom's magnetic moment ${\hat {\bf \mu}}$ to the states $a$, $a'$. Since the operator ${\hat {\bf \mu}}$  only acts on internal spin degrees of freedom of the electron and of the nucleus, it preserves both the principal quantum number and the orbital quantum number ${\rm L}$, and hence
\begin{eqnarray}
\!\!\!\!\!\!&\mu_i^{ab}&=\langle 1, 0 ,1/2;{\rm F}_a{\rm M}_a | {\hat \mu}_i | n_b {\rm L}_b {\rm J}_b;{\rm F}_b{\rm M}_b \rangle \nonumber \\
\!\!\!\!\!\!&=&\!\!\!\!\! \langle 1, 0 ,1/2; {\rm F}_a {\rm M}_a | {\hat \mu}_i | 1, 0,1/2;{\rm F}_b {\rm M}_b \rangle \; \delta_{n_b,1} \delta_{{\rm L}_b, 0},\label{matZee}
\end{eqnarray}
where ${\rm F}_a$ and ${\rm F}_b$ are ${\rm I} \pm 1/2$.
Eq. (\ref{matZee}) shows that the intermediate states $b$ contributing to the dynamic Zeeman shift  of  ground state hyperfine sub-levels are  the  same  ground state hyperfine levels.  For ${\rm L}_a=0$ the electron Land{\'e} factor is $g_{\rm J}=g=2$, and then
\be
\hat{\boldsymbol{\mu}}= -2 \mu_{\rm B}{\hat {\bf S}}+ g_{\rm I} \mu_{\rm n} {\hat {\bf I}} \;.
\ee
Since in a transition among  ground-state hyperfine Zeeman sub-levels it holds  $|E_a-E_b| \lesssim h \nu_{\rm HFS}$, it follows that at all temperatures  larger than the temperature of liquid helium  the ratio $(E_b-E_a)/k_B T$ is much smaller than one (as an example, in the case of hydrogen  $h \nu_{\rm HFS}/k_B T=0.017$ for $T=4$ K). It is therefore possible to use in Eq.~(\ref{farleyfor}) the  small $y$ asymptotic expansion of $F$ given in Eq.~(\ref{asF}), and then we get for the dynamic Zeeman shift  $\delta {\cal F}^{(0;\rm th|H)}_a$ the approximate formula,
\begin{eqnarray}
&&\delta {\cal F}^{(0;\rm th|H)}_a \simeq -\frac{8 \,\pi}{9\, \hbar\, c^3}\! \left(\frac{k_B T}{\hbar} \right)^2 \mu_{\rm B}^2 \nonumber \\ 
&&\times \sum_{i} \sum_{b \in \stackrel{\rm ground}{\rm \;states}} 
|S_i^{ab} |^2 (E_a-E_b)
\end{eqnarray}
By using this formula, one obtains the following estimate for the dynamic Zeeman shift $\delta \nu ^{(0;{\rm th}|\rm H)} _{{\rm I+1/2} \rightarrow {\rm I-1/2}}$  of the transition frequency between the states  ${\rm F=I+1/2}$ and  ${\rm F=I-1/2}$: 
\be 
\delta \nu ^{(0;{\rm th}|\rm H)}_{{\rm I+1/2} \rightarrow {\rm I-1/2}} \simeq -\frac{2 \,\pi}{9}\, \alpha_{\rm e} \left(\frac{k_B T}{m c^2} \right)^2 {\nu_{\rm HFS}}\;.\label{freeZee}
\ee
When applied to tritium, this formula gives  
\be
\delta \nu ^{(0; {\rm th} |\rm H)}_{{\rm I+1/2} \rightarrow {\rm I-1/2}}\simeq-2.0\times  \left(\frac{T}{T_{\rm room}}\right)^2 \,   10^{-8}\;{\rm Hz}\;.\label{freeZe}
\ee
The shifts obtained  for H and D are even smaller.
 
Now we  consider the dynamic Stark shift. For this, we need to consider the matrix elements of the electric dipole operator ${\hat {\bf d}}$. The selection rule $\Delta {\rm L}=\pm 1$ obeyed by  ${\hat {\bf d}}$ implies that only excited states with  ${\rm L}_b=1$ have  non-vanishing matrix elements with ground state levels,
\begin{eqnarray}
&d_i^{ab}&=\langle 1, 0, 1/2;{\rm F}_a{\rm M}_a | {\hat d}_i | n_b,{\rm L}_b,{\rm J}_b;{\rm F}_b{\rm  M}_b \rangle \nonumber \\
&=&\!\! \langle 1, 0 ,1/2;{\rm F}_a{\rm M}_a | {\hat d}_i | n_b, 1,{\rm J}_b; {\rm F}_b {\rm M}_b \rangle \;  \delta_{{\rm L}_b, 1}.\label{matStark}
\end{eqnarray}
The explicit expression of the matrix elements  $d_i^{ab}$ can be found in Appendix A. Eq.~(\ref{matStark}) implies that for all states $b$ such that $d_i^{ab} \neq 0$,  $E_b-E_a$ is of the order of the Bohr energy $w_0=13.6$ eV, and then   the ratio $(E_b-E_a)/k_B T$ is  large ($w_0/k_B T=526$ for $T=300$ K). Therefore, in Eq.~(\ref{farleyfor})  it is possible to use the asymptotic expansion of $F(y)$ for large $y$. This gives
\begin{eqnarray}
&&\delta {\cal F}^{(0;\rm th|E)}_a\simeq \frac{4 \,\pi^3 \hbar}{45\,  c^3}\! \left(\frac{k_B T}{\hbar} \right)^4  \sum_{i,b} 
\frac{|d_i^{ab} |^2} {E_a-E_b} \nonumber \\
&=&-\frac{2 \,\pi^3 \hbar}{45\,  c^3}\! \left(\frac{k_B T}{\hbar} \right)^4 {\rm Tr}\, \alpha^{(a)}(0) \;,
\end{eqnarray}
where in the last passage we used the formula for the electric polarizability Eq.~(\ref{defelpol}). By introducing the dimensionless polarizability ${\tilde \alpha}_{ij}^{(a)}= { \alpha}_{ij}^{(a)}/a_0^3$, where $a_0$ is the Bohr radius, we can recast the above formula as
\be
\delta {\cal F}^{(0;\rm th|E)}_a=-\frac{2 \,\pi^3 \hbar\,a_0^3}{45\,  c^3}\! \left(\frac{k_B T}{\hbar} \right)^4   {\rm Tr }\,\tilde{\alpha}^{(a)}(0)\;.\label{stark0}
\ee
The value of $ {\rm Tr }\,\tilde{\alpha}^{(a)}(0)$ for  ground-state hyperfine levels H, D and T atoms is very close to 13.5 .  
We thus get  for the dynamic Stark shift the estimate
\be
\frac{\delta {\cal F}^{(0;\rm th|E)}_a}{h} \simeq-\frac{\pi^2\,a_0^3}{45\,  c^3}\! \left(\frac{k_B T}{\hbar} \right)^4 \!\!\!13.5 = -\left(\frac{T}{T_{\rm room}}\right)^4\!\! \!0.039\;{\rm Hz}\;.
\ee
The shift of the hyperfine transition frequencies $\delta \nu^{(0;{\rm th}|\rm E)}_{aa'}$ is actually much smaller.  This is so because according to Eq.~(\ref{deltanu})  $\delta \nu^{(0;{\rm th}|\rm E)}_{aa'}$ actually  involves the {\it difference} $\Delta {\tilde \alpha}_{ij}^{(aa')}={\tilde \alpha}_{ij}^{(a)}-{\tilde \alpha}_{ij}^{(a')}$ among the polarizabilities of states $a$ and $a'$ 
\be
\delta \nu_{aa'}^{(0;{\rm th}|\rm E)}=-\frac{\pi^2 \,a_0^3}{45\,  c^3}\! \left(\frac{k_B T}{\hbar} \right)^4   \,{\rm Tr }\, \Delta \tilde{\alpha}^{(aa')}(0) \;.\label{freeSta}
\ee
It turns out that ground-state hyperfine sub-levels have almost identical electric polarizabilities, and therefore $ \Delta \tilde{\alpha}^{(aa')}(0)$ is very small.  For example, for the transition  ${\rm F=I+1/2} \rightarrow {\rm F=I-1/2}$, we find that  for H, D and  T  the value of ${\rm Tr }\, \Delta \tilde{\alpha}^{(aa')}(0)$  is equal to $6.9\times 10^{-6}$,   $ 1.6\times 10^{-6}$  and  $  7.3 \times 10^{-6}$,  respectively.  This implies that in the case of T the frequency shift has the extremely small value of:
\be
\delta \nu^{(0;{\rm th}|\rm E)}_{{\rm I+1/2} \rightarrow {\rm I-1/2}}=- 2.1 \times \left(\frac{T}{T_{\rm room}}\right)^4 10^{-8}  \;{\rm Hz}\;.\label{freeSt}
\ee
In the case of H and D the obtained shifts are even smaller.
According to Eq.~(\ref{totalshift0}) the total shift $\delta \nu^{(0;{\rm th})}_{aa'}$ is  the sum of the dynamic Zeeman shift $\delta \nu^{(0;{\rm th}|\rm H)}_{aa'}$ in Eq.~(\ref{freeZee}) and  the dynamic Stark shift $\delta \nu^{(0;{\rm th}|\rm E)}_{aa'}$ in  Eq.~(\ref{freeSta}). 
It is clear from Eqs.~(\ref{freeZe}) and (\ref{freeSt}) that the resulting shift is too small to be measured, for practically attainable temperatures.

Finally, we consider the radiative width $\Delta \omega_{1/2}$ of the transition. Using Eq.~(\ref{width}) we find that $\Delta \omega_{1/2}$ is extremely small in free-space.  For $T=300$ K, $\Delta \omega_{1/2}=2.5 \times 10^{-11}$ Hz for H,   $\Delta \omega_{1/2}=2.7 \times 10^{-12}$ Hz for D,  and $\Delta \omega_{1/2}=2.9 \times 10^{-11} $ Hz for T.
The conclusion of the above findings is that the shifts $\delta \nu$ and the widths $\Delta \omega_{1/2}$ of hyperfine transitions in free space are both too  small to be measurable.


\section{Atom in a planar cavity}

\begin{figure}[ht]
\includegraphics [width=.9\columnwidth]{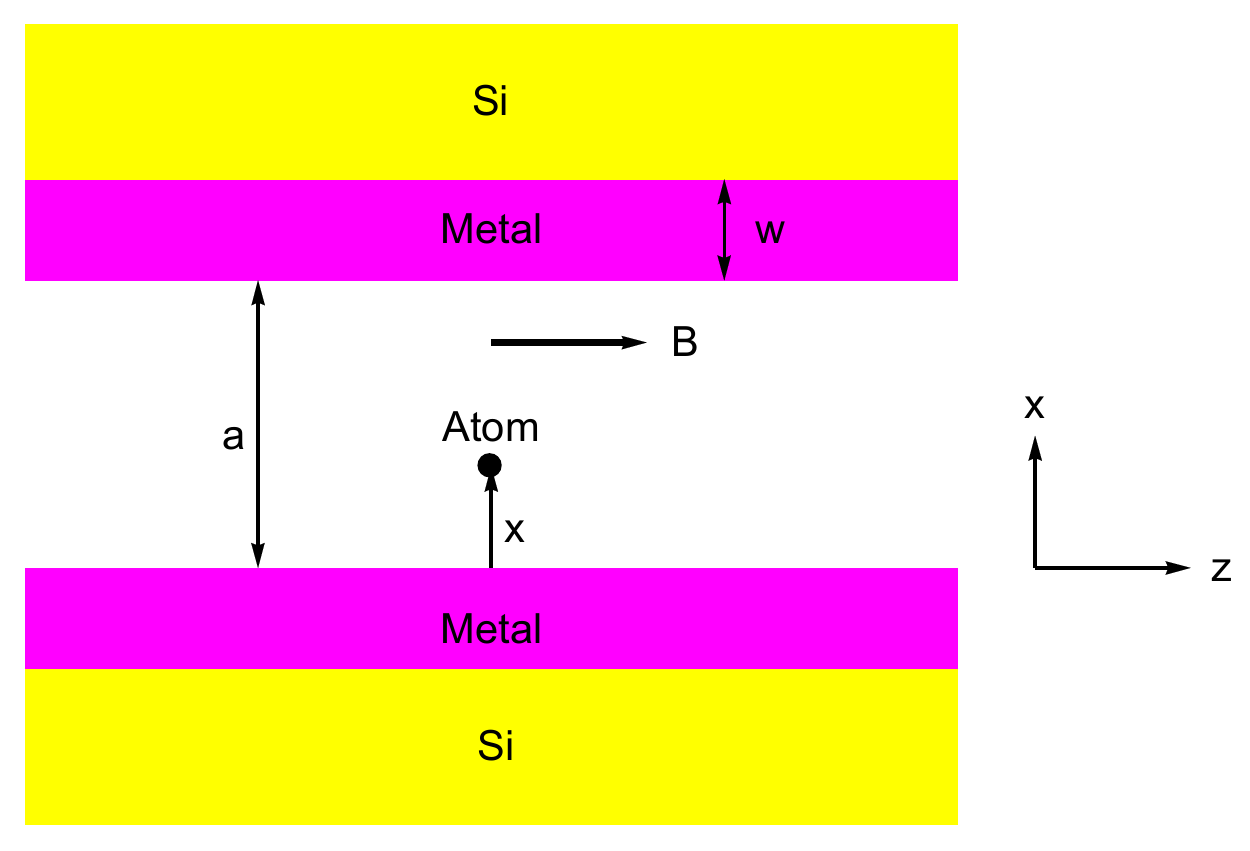}
\caption{\label{Fig3}  Planar cavity of width $a$, consisting of two layered plane-parallel metallic mirrors.  The atom is at distance $x$ from the lower mirror. The uniform magnetic field $B$ is directed along the $z$ direction.}
\end{figure}

In this Section we compute the energy shifts and the widths of the transitions between hyperfine ground-state Zeeman sub-levels of an atom placed inside a plane-parallel metallic cavity of width $a$, constituted by two identically constructed 2-layer mirrors consisting of a metallic layer of thickness $w$, deposited on top of a thick Si substrate. The cavity, which is in equilibrium with the environment at temperature $T$, is embedded in a uniform magnetic field $B$, whose direction is parallel to the surface of the mirrors. The $z$ axis coincides with the direction of the $B$-field, while the $x$-axis is perpendicular to the mirrors surfaces. $x$ is the distance of the atom from the surface of the lower mirror (see Fig.~\ref{Fig3}).

We shall study in detail the hyperfine transition $({\rm F=1,m_F=0}) \rightarrow ({\rm F=0,m_F=0})$ of H and T atoms,  immersed in a weak magnetic field of about $B=0.01$G. The reason behind  this choice is explained as follows. The computations presented below  will show that the  frequency shift induced by the CP interaction is  rather small (of the order of a few tens of Hz). This implies that in order to have a chance for observing this little effect, it is necessary to isolate a transition which is robust against the perturbing effect of small  inhomogeneities and/or uncertainties of external magnetic fields.   The  transition $({\rm F=1,m_F=0}) \rightarrow ({\rm F=0,m_F=0})$  satisfies precisely this requirement. In Sec.~III, we indeed showed that in a weak field $B \ll A_{\rm J}$ the Zeeman effect shifts the energy of the hyperfine state $(\rm F,M_F)$ by an amount proportional to $\rm M_F$ (see Eq.~(\ref{enZee})). This implies that to first order in $B$, states with $\rm M_F=0$  are immune to the Zeeman shift, while states with ${\rm M}_F \neq 0$  get shifted.  This feature   implies that by placing the cavity in a weak $B$-field, we can   separate   the  Zeeman-sensitive transitions  $({\rm F=1,m_F=\pm 1}) \rightarrow ({\rm F=0,m_F=0})$ from the  immune transition, without appreciably perturbing the frequency of the latter.

To get a quantitative insight of the magnitude of the external field $B$ that  achieves this goal, consider the example of a $B$-field of 0.01G. Using Eq.~(\ref{enZee}) one finds that  the energy of the state $(\rm F=1,M_F=1)$  gets shifted upwards by 14 kHz, while the energy of the state $(\rm F=1,M_F=-1)$  gets shifted downwards by  the same amount. Thus,  a weak $B$-field  of 0.01G is sufficient to separate the  stable transition $({\rm F=1,m_F=0}) \rightarrow ({\rm F=0,m_F=0})$ from both   transitions  involving the ${\rm M_F=\pm1}$ states.  On the other hand, a  $B$-field of 0.01G has a negligible effect on the frequency of the transition $({\rm F=1,m_F=0}) \rightarrow ({\rm F=0,m_F=0})$.  In fact, an evaluation of the exact Breit-Rabi formula Eq. (\ref{BR}) shows    that the upper state $(\rm F=1,M_F=0)$ is shifted upwards by 0.14 Hz, while the lower state   $(\rm F=0,M_F=0)$ is shifted downwards by -0.14 Hz. This implies that the $B$-field shifts the  frequency of the $({\rm F=1,m_F=0}) \rightarrow ({\rm F=0,m_F=0})$ transition by just 0.28 Hz, which is much smaller than the CP shift  we obtain for this transition (see Figs.~3 and 5 below). After these general considerations, we  turn to the computation of the shift $\delta \nu$.

 As shown in Sec.~II, the shift $\delta \nu_{aa'}$ of the frequency of the transition $a \rightarrow a'$  is proportional to the difference between the free energy shifts $\delta {\cal F}_a$ and $\delta {\cal F}_{a'}$ of the two states. The   shifts $\delta {\cal F}_a$ for an atom placed in a cavity can be computed by  using Eqs.~(\ref{splitdeltaF}-\ref{deltaFsc1}), once we substitute in those equations the scattering Green functions of the cavity.  The explicit formulae of the these Green functions are provided in Appendix B.  As it can be seen from Eqs.~(\ref{Exx}) and (\ref{Ezz}) the  Green functions involve the reflection coefficients of the mirrors $R^{(k)}_{\alpha}$, which in turn involve the complex frequency-dependent permittivities $\epsilon_a(\omega)$ of the materials constituting the mirrors (see Eqs.~(\ref{refl1}-\ref{freTM})). For the permittivities we use the following models.  
For the metals, we used the simple Drude model
 \be
\epsilon_{\rm Dr}(\omega)=1-\frac{\Omega_p^2}{\omega(\omega+{\rm i} \gamma)}\;,
 \ee 
where $\Omega_p$ and $\gamma$ are, respectively, the plasma and the relaxation frequencies.  This simple model 
is adequate for our problem, because it turns out that up-to negligible terms, the  shifts $\delta {\cal F}_a$  are determined by the resonant contribution to the dynamic Zeeman effect, which according to Eq.~(\ref{deltaFsc1}) involves the values of the Green tensors at the frequencies of the hyperfine transitions. The latter frequencies belong to the GHz region, where the Drude model provides an accurate description of the  electromagnetic response of metals.   

We recall that the plasma frequency $\Omega_p$ is independent of  temperature, while the relaxation frequency is temperature dependent \cite{kittel}. It is well known \cite{kittel} that  in the range of temperatures  $T_{\rm D}/4 < T < T_{\rm room} = 300$K, where $T_{\rm D}$ is the Debye temperature, the relaxation frequency  depends linearly on the  temperature,
\be
\gamma=\gamma_{\rm room} [1+(T-T_{\rm room})\alpha]\;.
\ee 
At lower temperatures,   $\gamma$   reaches a saturation value $\gamma_s=\gamma_{\rm room}/{\rm RRR}$ where the  ${\rm RRR}$ ratio is determined by the amount of impurities, which depends on  sample preparation. In ultra-clean samples, and for temperatures extending from $T_D/4$ down to liquid helium temperature,  $\gamma(T)$   decreases like $T^5$, in accordance with the Bloch-Gr\"uneisen law \cite{ash}. At even lower temperatures $\gamma(T) \sim T^2$ for metals with perfect crystal lattices \cite{kittel}.
In Table II we list the values of the Drude parameters, Debye temperature and the thermal coefficient $\alpha$ for some metals. In the numerical computations for $T=$70 K presented below, we set $\gamma=\gamma_{\rm room}/{\rm RRR}$ for all considered metals.
\begin{table}[h]
\begin{tabular}{ccccc} \hline 
Metal &\;\;\; $\Omega_p$(eV/$\hbar$) &\;\;\; $\gamma_{\rm room}$ (eV/$\hbar$) & \;\;\; $\alpha \times 10^3({\rm K}^{-1})$ &  \;\; $T_{\rm D}$(K)  \\ \hline \hline
$\!\!{\rm Au}$ &\;9\;\;  &  0.035       & 3.4  & 165   \\  \hline
 $\!\! {\rm Al}$ & \;11.5  &   0.050      & 4.3 & 428 \\  \hline
 $\!\!{\rm Ag}$ & 9.014 &   0.018 & 4.0  & 225  \\  \hline
  $\!\!{\rm Pt}$ & 4.89 &   0.07 & 3.9 & 240   \\
 \hline 
\end{tabular}
\caption{Values of the Drude parameters \cite{ordal}, Debye temperatures $T_{\rm D}$ \cite{kittel} and thermal coefficients $\alpha$ for some metals.}
\label{tab.2}
\end{table}

For the permittivity of the Si substrate  we used the formula \cite{kittel}
\be
\epsilon_{\rm Si}(\omega)=\epsilon_{\infty} + \frac{\omega_{\rm UV}^2}{\omega_{\rm UV}^2-\omega^2-{\rm i} \omega \gamma_{\rm Si}}  (\epsilon_0-\epsilon_{\infty})\;,
\ee  
with $\epsilon_{\infty} =1.035$, $\epsilon_0=11.67$, $\omega_{\rm UV}=6.6 \times 10^{15}$ rad/s, and $\gamma_{\rm Si}=1.52 \times 10^{12}$ rad/s.

We are now ready to compute  the shift $\delta \nu_{aa'}$  of the hyperfine transition of an atom placed inside the cavity.  According to Eq.~(\ref{splitdeltaF}) the shift $\delta \nu_{ab}$ is the sum of the free-space shift $\delta \nu^{(0;{\rm th})}_{aa'}$ and a scattering contribution $\delta \nu^{({\rm sc})}_{aa'}$,
\be
\delta \nu_{aa'}=\delta \nu^{(0;{\rm th})}_{aa'}+\delta \nu^{({\rm sc})}_{aa'}\;.
\ee
The free-space shift $\delta \nu^{(0;{\rm th})}_{aa'}$ was computed in the previous Section and was found to be negligible.  In view of Eqs.~(\ref{absE}-\ref{deltaFsc1}) the scattering contribution $\delta \nu^{({\rm sc})}_{aa'}$ to the shift  is the sum of six terms which are written as
\begin{eqnarray}
\delta \nu^{({\rm sc})}_{aa'}&=&\delta \nu^{({\rm abs|E})}_{aa'}+\delta \nu^{({\rm em|E})}_{aa'}+\delta \nu^{({\rm non\;res|E})}_{aa'}\nonumber \\
&+& \delta \nu^{({\rm abs|H})}_{aa'}+\delta \nu^{({\rm em|H})}_{aa'}+\delta \nu^{({\rm non\;res|H})}_{aa'}\;.
\end{eqnarray}
Since we are interested in the shifts of ground-state hyperfine transitions, both states $a$ and $a'$ are ground-state hyperfine sub-levels. This implies at once that  
\be
\delta \nu^{({\rm em|E})}_{aa'}=0\;.
\ee
The above identity is a consequence of the fact that  for any ground-state level $a$ the shift $\delta {\cal F}_a^{(\rm em|E)}$ in Eq.~(\ref{emE})  vanishes identically.   This is so because the intermediate states $b$ that contribute to $\delta {\cal F}_a^{(\rm em|E)}$ must have a lower energy than $a$, and therefore they  are ground-state hyperfine levels.  However the matrix elements of the electric-dipole operators between two ground-state levels are zero, and therefore  $\delta {\cal F}_a^{(\rm em|E)}$ vanishes.  Next, we consider the contribution $\delta \nu^{({\rm abs|E})}_{aa'}$. This term is negligible, since according to Eq.~(\ref{absE}) it only involves intermediate states $b$ that are coupled to the ground-state hyperfine sub-levels $a$ and $a'$ by the electric-dipole operator. Because of the selection rule $\Delta {\rm L}=\pm1$, these intermediate states  are excited states with ${\rm L}=1$, and then for the temperatures that we consider, the  Bose factor $[\exp(\hbar \omega_{ab}/k_B T)-1]^{-1} \simeq [\exp(w_0/k_B T)-1]^{-1} \ll 1$ leads to a strong suppression of $\delta \nu^{({\rm abs|E})}_{aa'}$.  Next, we consider $\delta \nu^{({\rm non\;res|E})}_{aa'}$. According to Eq.~(\ref{nonresE}), this contribution can be recast in the form
\be
\delta \nu^{({\rm non\;res|E})}_{aa'}\!\!= -k_B T a_0^3 \sum_{n=0}^{\infty}\;\!\!'  {\cal E}^{(\rm sc)}_{i j}({\bf r}_0, {\bf r}_0; {\rm i} \,\xi_n )\Delta {\tilde \alpha}^{(a a')}_{ij}({\rm i} \,\xi_n) \;.
\ee 
We pointed out earlier that the electric polarizabilities ${\tilde \alpha}^{(a)}_{ij}$ of ground-state hyperfine  sub-levels are almost identical to each other. Because of this feature  the shift $\delta \nu^{({\rm non\;res|E})}_{aa'}$ is very small.  As an example,  consider the transition $({\rm F=1,m_F=0}) \rightarrow ({\rm F=0,m_F=0})$  of an H atom placed  in a Au cavity at room temperature, having a width $a$ of one micron, in zero magnetic field. If the atom is placed at the center of the cavity (i.e. for $x=a/2$), we find  $\delta \nu^{({\rm non\;res|E})}_{1,0;0,0}=-1.3 \times 10^{-4}$ Hz.  If the atom is moved at a distance $x=200$  nm from the lower mirror,  we find  $\delta \nu^{({\rm non\;res|E})}_{1,0;0,0}=-2.4 \times 10^{-3}$ Hz. Shifts of a comparably small magnitude are obtained also for the other transitions, as well as for D and T. The above considerations show that the dynamic Stark shift of the ground-state hyperfine transition frequencies is negligibly small. 

We consider now the dynamic Zeeman shift. It turns out that the non-resonant contribution $\delta \nu^{({\rm non\;res|H})}_{aa'}$ is always very small, compared to the resonant contribution $\delta \nu^{({\rm res|H})}_{aa'}=\delta \nu^{({\rm abs|H})}_{aa'}+\delta \nu^{({\rm em|H})}_{aa'}$. Consider again  the transition $({\rm F=1,m_F=0}) \rightarrow ({\rm F=0,m_F=0})$  of an H atom placed in a Au cavity, under the same conditions considered above. At room temperature,  the non-resonant shift  for $x=a/2$   is  $\delta \nu^{({\rm non\;res|H})}_{aa'}=-3.14 \times 10^{-6}$ Hz, to be contrasted with $\delta \nu^{({\rm res|H})}_{aa'}=7.44$ Hz.  Similarly, for $x=200$ nm we find $\delta \nu^{({\rm non\;res|H})}_{aa'}=-2.15 \times 10^{-5}$ Hz, while $\delta \nu^{({\rm res|H})}_{aa'}=8.31$ Hz. For $T=70$ K, and for $x=a/2$   we find $\delta \nu^{({\rm non\;res|H})}_{aa'}=-1.6 \times 10^{-5}$ Hz , to be contrasted with  $\delta \nu^{({\rm res|H})}_{aa'}=10.12$ Hz,   while for $x=200$ nm we find $\delta \nu^{({\rm non\;res|H})}_{aa'}=-9.8 \times 10^{-5}$ Hz, and $\delta \nu^{({\rm res|H})}_{aa'}=12.9$ Hz. From this we  conclude that, up to negligible corrections, the shift $\delta \nu_{aa'}$ of the hyperfine ground-state transitions is entirely due to the resonant contributions to the dynamic Zeeman effect, i.e.,
\be
\delta \nu_{aa'}\simeq  \delta \nu^{({\rm res|H})}_{aa'} \equiv \delta \nu^{({\rm abs|H})}_{aa'}+\delta \nu^{({\rm em|H})}_{aa'}\;.
\ee
An important feature of the above result is its {\it robustness}. Since the dynamic Zeeman shifts of the free-energies  $\delta {\cal F}_a^{(\rm abs|H)}$ and $\delta {\cal F}_a^{(\rm em|H)}$ depend only on the hyperfine transition frequencies $\omega_{ab}$ (see Eqs.~(\ref{absH})) and (\ref{emH})), it is clear that the frequency shift  $\delta \nu_{aa'}$ is independent of the value of the constant $E_{\rm J}$ in Eq.~(\ref{EFnoB}), whose effect is to  just shift the  energies of the hyperfine levels by an irrelevant overall constant,  which does not change the hyperfine transition frequencies $\omega_{ab}$.  This implies that the inclusion in  $E_{\rm J}$ of higher order corrections in $\alpha_{\rm e}$, like for example for the Lamb shift, is irrelevant as far as they are independent of the total angular momentum ${\rm F}$.  The transition frequencies $\omega_{ab}$ are determined  solely by the experimentally known frequency $\nu_{\rm HFS}$ and by the magnetic field $B$, in accordance with the Breit-Rabi formula in Eq.~(\ref{BR}).

\begin{figure}[ht]
\includegraphics [width=.9\columnwidth]{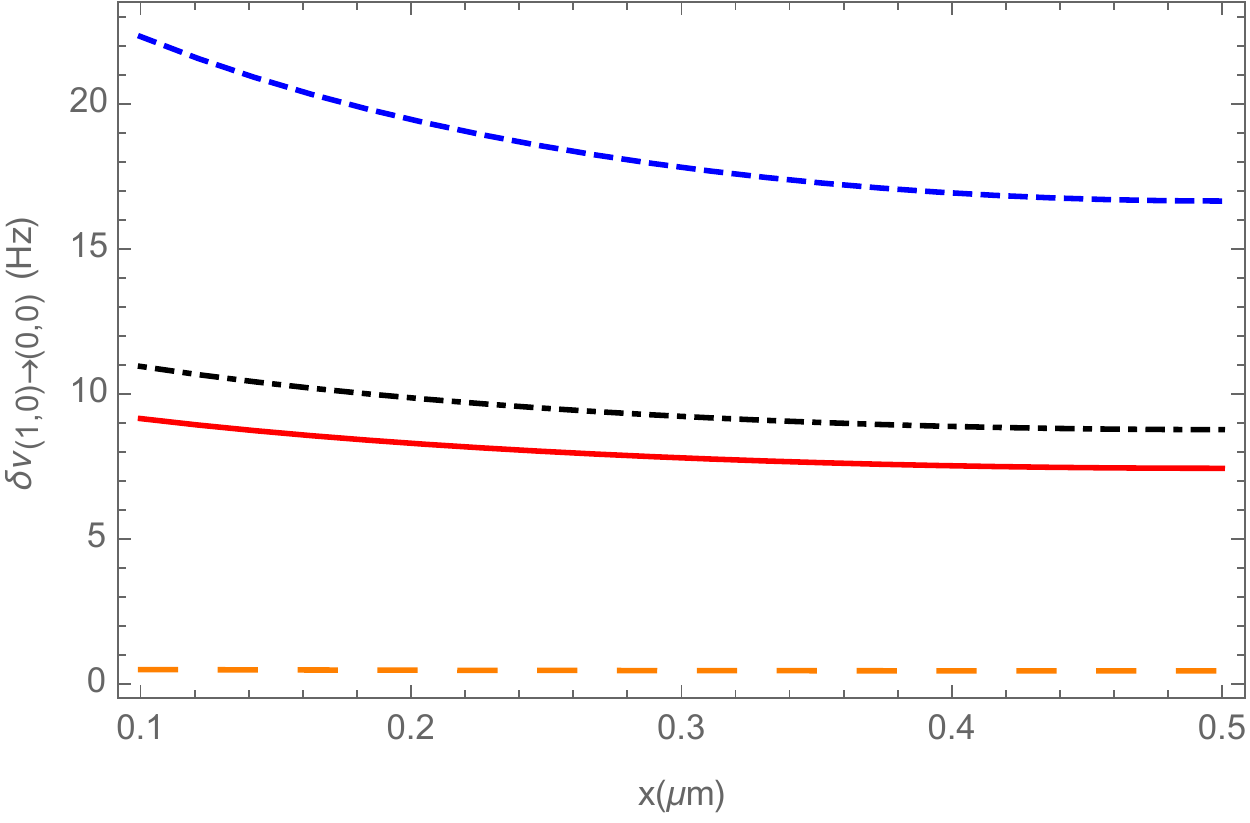}
\caption{\label{shift300}  Shift $\delta \nu$ (in Hz) of the H hyperfine transition $({\rm F=1,m_F=0}) \rightarrow ({\rm F=0,m_F=0})$ (in Hz), versus the minimum atom-mirror separation $x$ (in micron). The atom is placed inside a   metallic cavity of width $a=1$ micron at room temperature, in a magnetic field $B$=0.01G. The thickness of the metallic layer is 5 micron. The four curves, from top to bottom, are for a cavity made of Ag, Al, Au and Pt, respectively.}
\end{figure}
\begin{figure}[ht]
\includegraphics [width=.9\columnwidth]{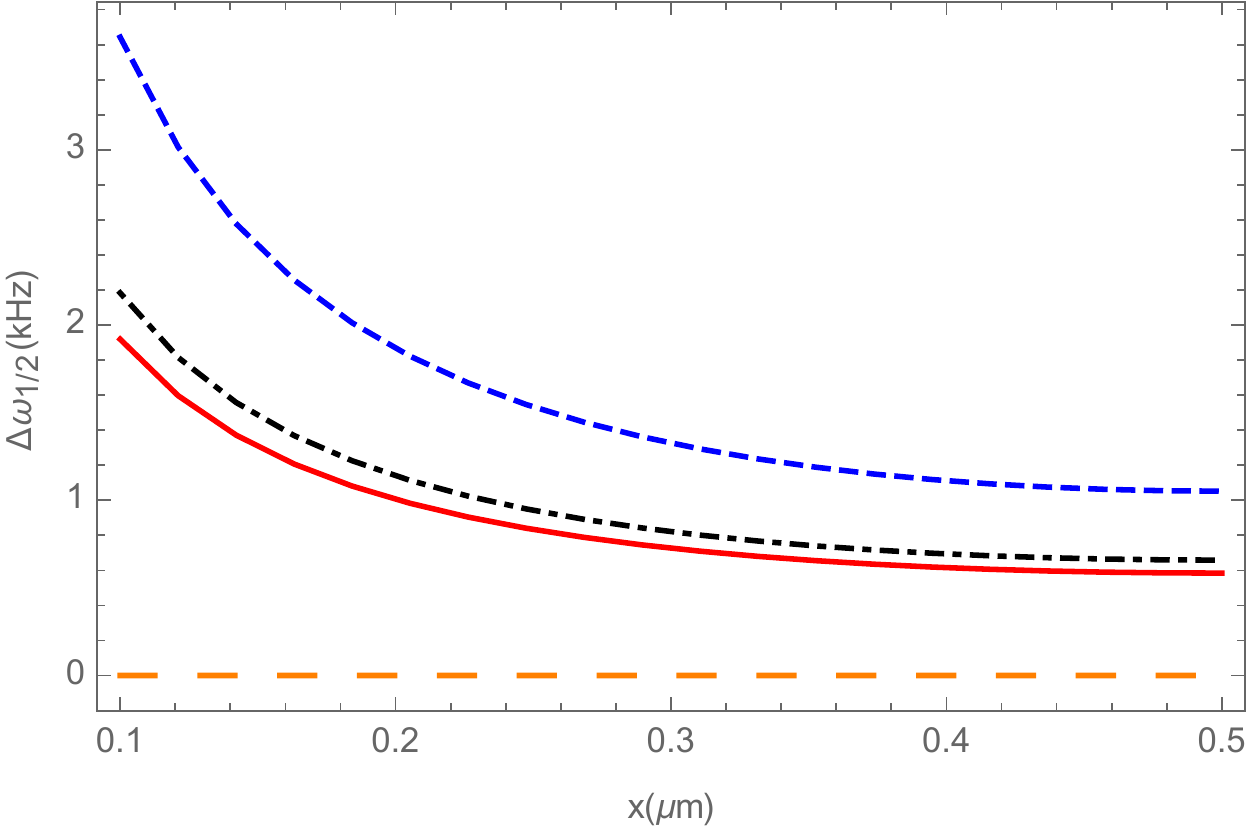}
\caption{\label{rate300}   Half-width $\Delta \omega_{1/2}$ (in kHz) of the H hyperfine transition $({\rm F=1,m_F=0}) \rightarrow ({\rm F=0,m_F=0})$ (in Hz), versus the minimum atom-mirror separation $x$ (in micron). The atom is placed inside a metallic cavity of width $a=1$ micron at room temperature, in a magnetic field $B$=0.01G. The  metallic layers have a thickness  $w$=5 micron.  The four curves, from top to bottom, are for a cavity made of Ag, Al, Au and Pt, respectively.}
\end{figure}
\begin{figure}[ht]
\includegraphics [width=.9\columnwidth]{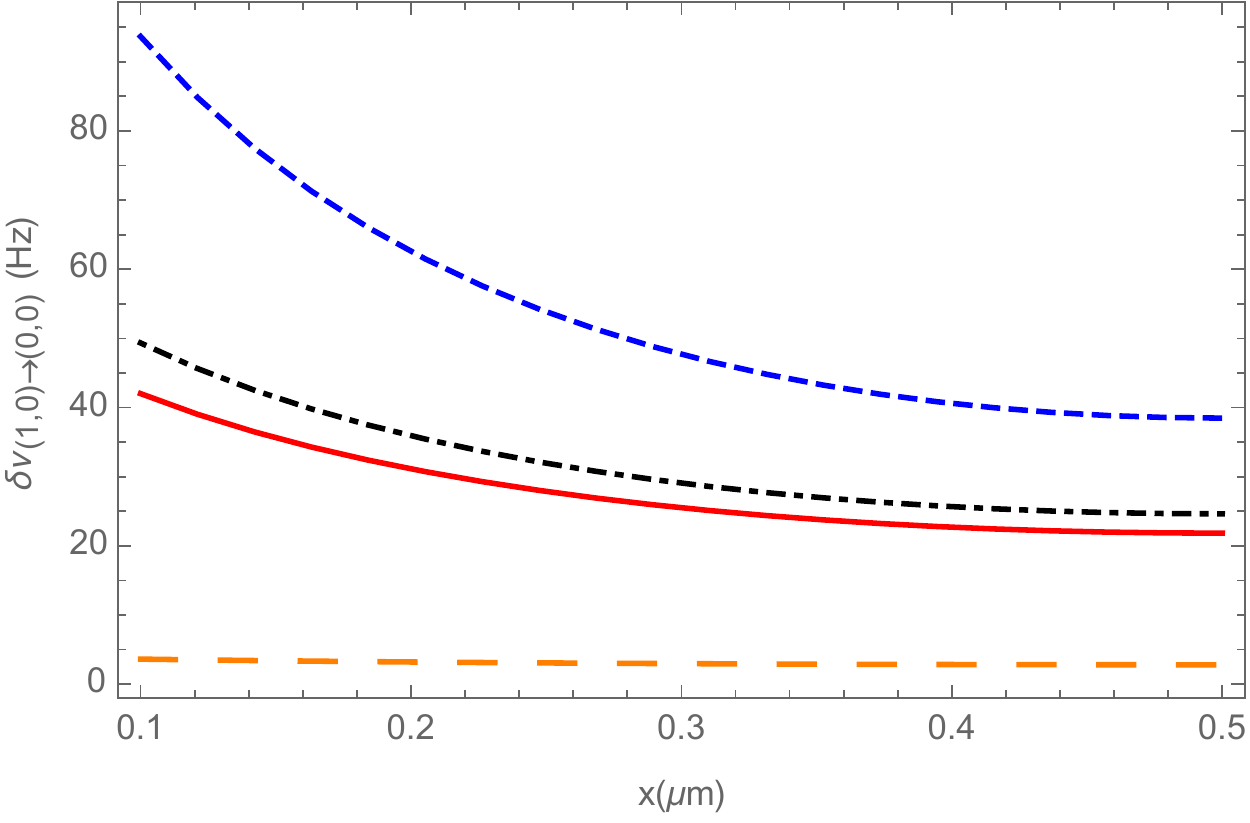}
\caption{\label{shiftH70}  Shift $\delta \nu$ (in Hz) of the H hyperfine transition $({\rm F=1,m_F=0}) \rightarrow ({\rm F=0,m_F=0})$ (in Hz), versus the minimum atom-mirror separation $x$ (in micron). The atom is placed inside a   metallic cavity of width $a=1$ micron at a temperature $T=$ 70 K (RRR=10), in a magnetic field $B$=0.01G. The  metallic layers have a thickness  $w$=5 micron. The four curves, from top to bottom, are for a cavity made of Ag, Al, Au and Pt, respectively.}
\end{figure}
\begin{figure}[ht]
\includegraphics [width=.9\columnwidth]{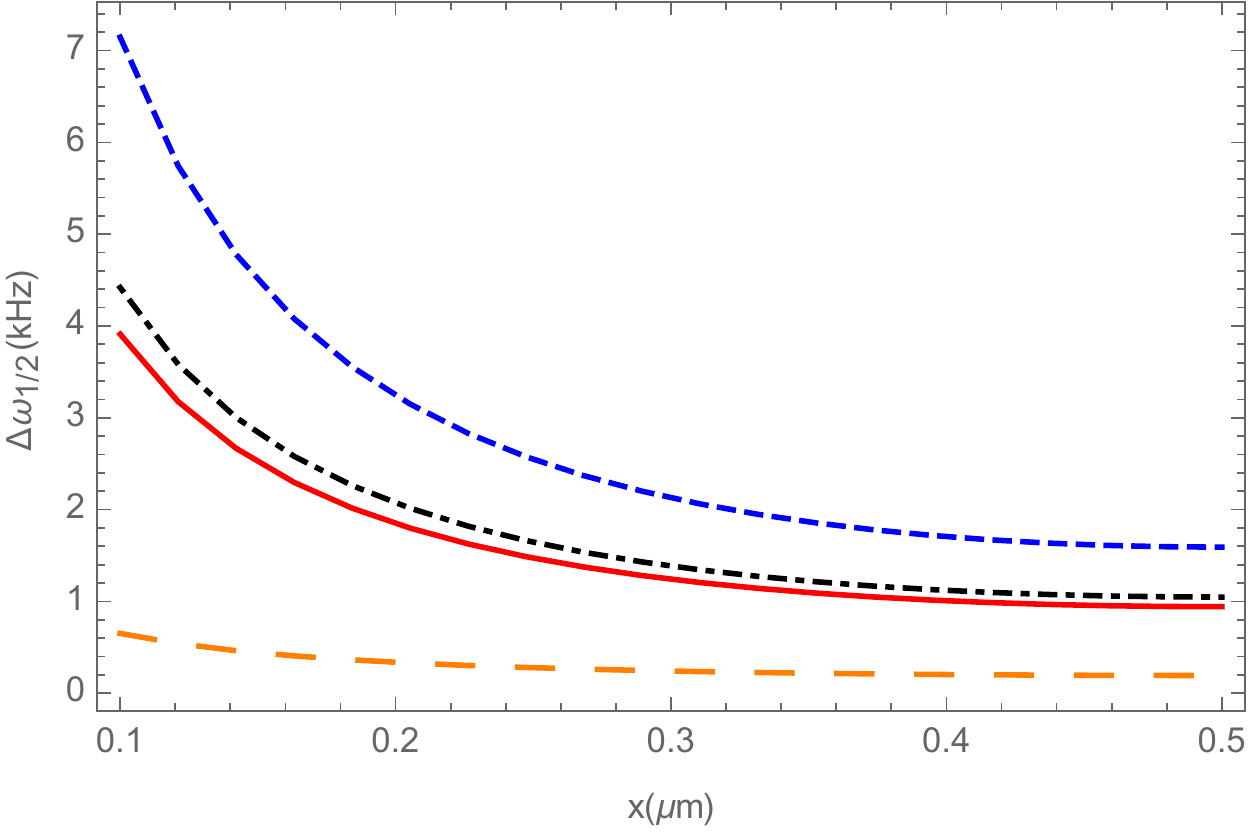}
\caption{\label{rateH70}   Half-width $\Delta \omega_{1/2}$ (in kHz) of the H hyperfine transition $({\rm F=1,m_F=0}) \rightarrow ({\rm F=0,m_F=0})$ (in Hz), versus the minimum atom-mirror separation $x$ (in micron). The atom is placed inside a  metallic cavity of width $a=1$ micron at a temperature $T=$ 70 K (RRR=10), in a magnetic field $B$=0.01G. The  metallic layers have a thickness  $w$=5 micron. The four curves, from top to bottom, are for a cavity made of Ag, Al, Au and Pt, respectively.}
\end{figure}

\begin{figure}[ht]
\includegraphics [width=.9\columnwidth]{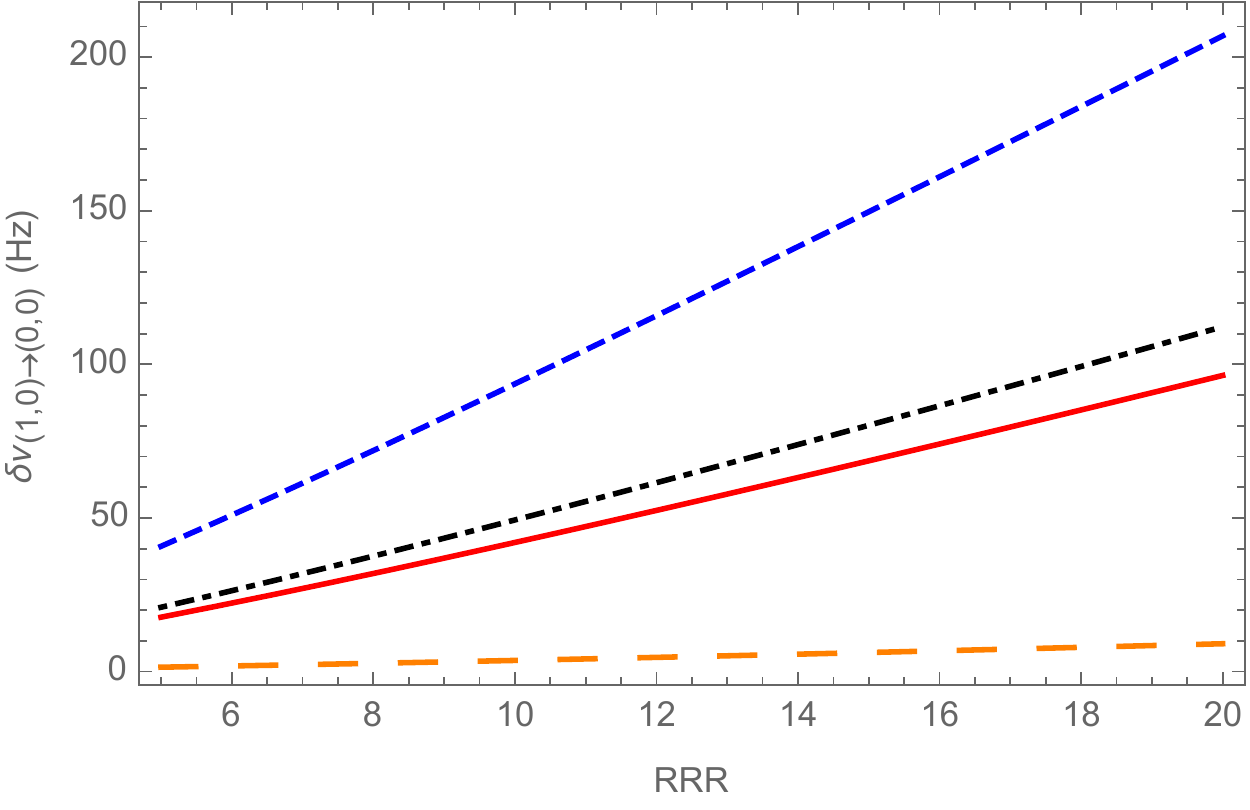}
\caption{\label{shiftH7RRR0}  Shift $\delta \nu$ (in Hz) of the H hyperfine transition $({\rm F=1,m_F=0}) \rightarrow ({\rm F=0,m_F=0})$ (in Hz), versus  the RRR ratio of the mirrors. The atom is placed at  a distance  $x$=100 nm from one of the mirrors of a   metallic cavity of width $a=1$ micron at a temperature $T=$ 70 K, in a magnetic field $B$=0.01G. The  metallic layers have a thickness  $w$=5 micron. The four curves, from top to bottom, are for a cavity made of Ag, Al, Au and Pt, respectively.}
\end{figure}

\begin{figure}[ht]
\includegraphics [width=.9\columnwidth]{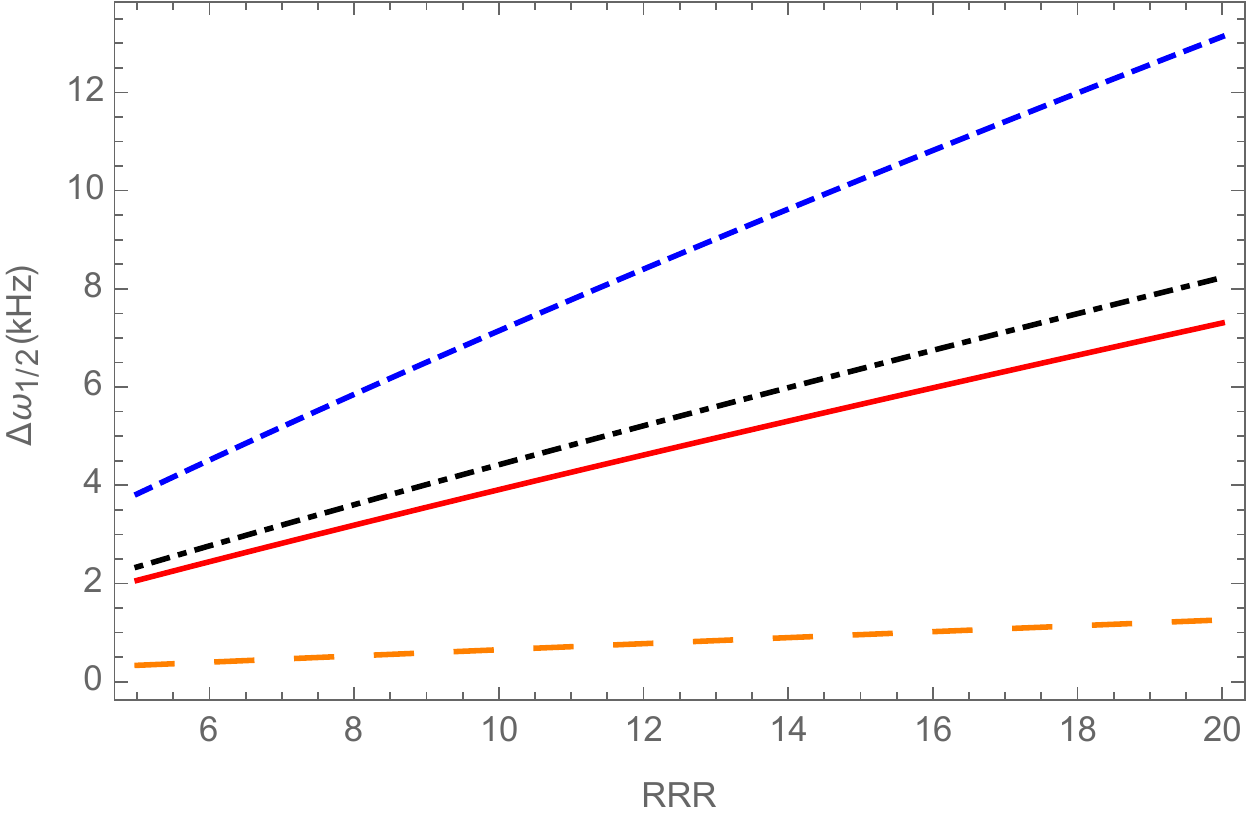}
\caption{\label{rateH70RRR}   Half-width $\Delta \omega_{1/2}$ (in kHz) of the H hyperfine transition $({\rm F=1,m_F=0}) \rightarrow ({\rm F=0,m_F=0})$ (in Hz), versus the RRR ratio of the mirrors. The atom is placed   at  a distance  $x$=100 nm from one of the mirrors of a  metallic cavity of width $a=1$ micron at  temperature $T=$ 70 K, in a magnetic field $B$=0.01G. The  metallic layers have a thickness  $w$=5 micron. The four curves, from top to bottom, are for a cavity made of Ag, Al, Au and Pt, respectively.}
\end{figure}

The results of our numerical computations are displayed in Figs.~3 to 9. In Fig.~\ref{shift300} and in Fig.~\ref{shiftH70} we show the shift  $\delta \nu$ (in Hz) versus separation $x$ (in micron) of the hyperfine transition $({\rm F=1,m_F=0}) \rightarrow ({\rm F=0,m_F=0})$  of an H atom placed in a metallic cavity having a width $a$ of one micron, in a magnetic field $B$=0.01G. The temperature of the cavity is $T=300 K$ in Fig.~\ref{shift300}, and $T=70 K$ (assuming an RRR ratio of 10)  in Fig.~\ref{shiftH70}. The four displayed curves, from top to bottom, are for a cavity made of Ag, Al, Au and Pt, respectively.  The  metallic layers have a thickness $w$=5 micron. In Fig.~\ref{rate300} and in Fig.~\ref{rateH70} we display the respective half-widths $\Delta \omega_{1/2}$ (in kHz).   Comparison of Fig.~\ref{shiftH70} with Fig.~\ref{shift300}  shows that the  magnitude of the shift $\delta \nu$ increases  as the temperature of the cavity is decreased. This  behavior is explained by  the fact that  the conductivity of the mirrors becomes larger at lower temperatures.  This  is further  demonstrated by Fig.~(\ref{shiftH7RRR0}) which displays the shift for an H atom at a distance of 100nm from one of the mirrors, as a function of the RRR ratio of the mirrors, for $T=70$ K. Fig.~\ref{rateH70RRR} displays the corresponding behavior for the rate, which also increases with RRR.   The results obtained for T are not shown here as they are similar to those for H. 


\begin{figure}[ht]
\includegraphics [width=.9\columnwidth]{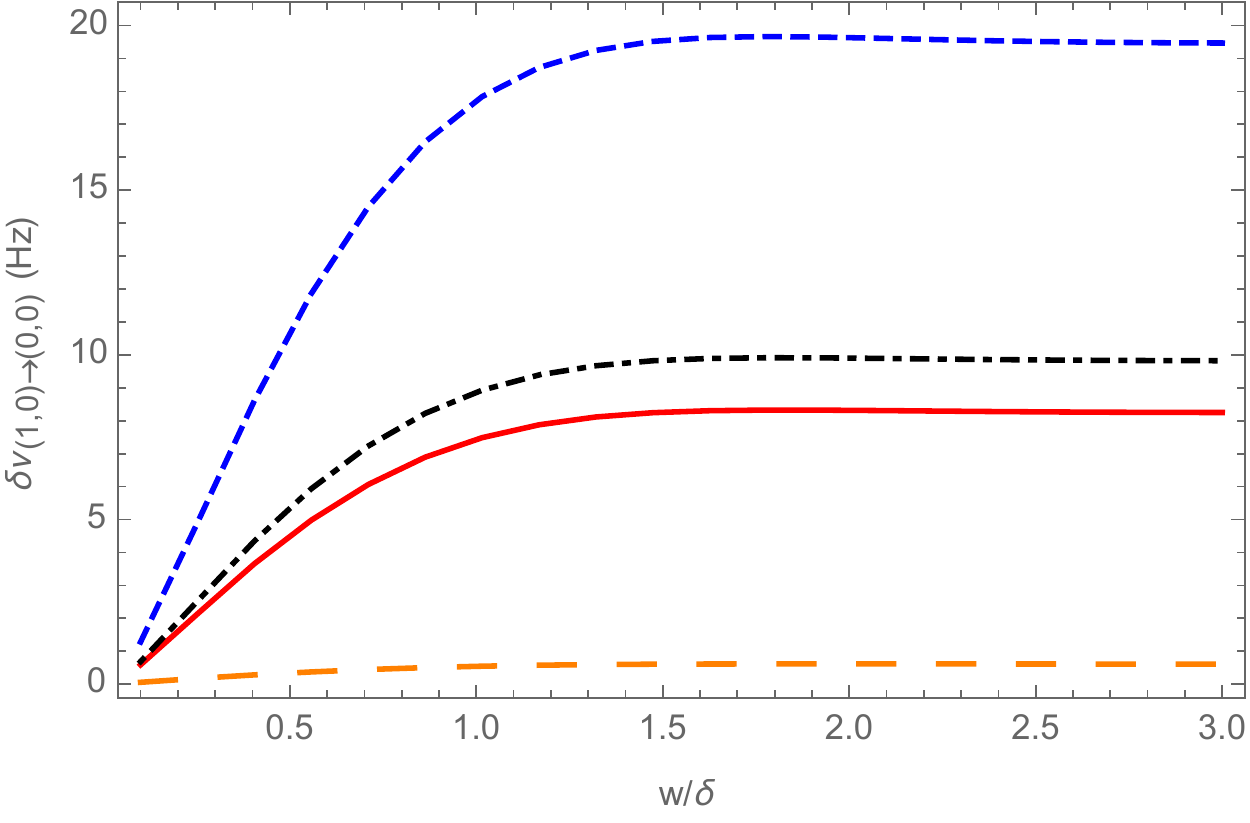}
\caption{\label{skin}  Shift $\delta \nu$ (in Hz) of the hyperfine transition $({\rm F=1,m_F=0}) \rightarrow ({\rm F=0,m_F=0})$ of an H atom placed at a distance $x=200$ nm from one of the mirrors of a  metallic cavity of width $a=1$ micron at room temperature, versus the thickness $w$ of the metallic layer (in units of the skin depth $\delta$ corresponding to the frequency $\nu_{\rm HFS}$). The external magnetic field is $B$=0.01G.   The four curves, from top to bottom, are for a cavity made of Ag, Al, Au and Pt, respectively.}
\end{figure}

In Fig.~\ref{skin} we finally  plot the shift $\delta \nu$ (in Hz) of the hyperfine transition $({\rm F=1,m_F=0}) \rightarrow ({\rm F=0,m_F=0})$ an H atom placed at a distance $x=200$ nm from one of the mirrors of a  metallic cavity of width $a=1$ micron at room temperature, versus the thickness $w$ of the metallic layer, divided by the skin depth $\delta=c/2\pi \sqrt{\nu_{\rm HFS} \,\sigma}$ corresponding to the frequency $\nu_{\rm HFS}$. The four curves, from top to bottom, are for a cavity made of Ag, Al, Au and Pt, respectively. The respective values of the skin depths are $\delta=1.7\;\mu$m,  $2.2\;\mu$m, $2.4\;\mu$m and
$6.2\;\mu$m. The plot shows that for a thickness $w$ of the metallic layer approximately equal to twice the skin depth the obtained shift $\delta \nu$ is indistinguishable from that of an infinitely thick mirror.  

\section{Conclusions}
 
The Casimir-Polder interaction of an atom prepared in a well-defined energy state may deviate strongly from the  Casimir-Polder energy predicted by Lifshitz theory for a fully thermalized atom, as a result of resonant  virtual-photon absorption and emission processes, that are absent when the atom is in a thermal  equilibrium  state. This  non-equilibrium  interaction can be measured by spectroscopic means, by observing frequency shifts suffered by atoms inside a cavity. In this paper we  computed  the frequency shift of ground-state hyperfine transitions of an  H atom placed in a metallic cavity at finite temperature.  We found that the resonant Casimir-Polder  interaction of the atom's magnetic moment with the fluctuating magnetic field existing in the cavity causes  a significant shift of hyperfine transitions, and it also leads to a very large increase of their widths,  as compared to an atom in free-space exposed to black-body radiation at the same temperature as the cavity.  By considering cavities made of different metals and hold at different temperatures, we established that larger frequency shifts   are obtained in cavities made of  metals having a large conductivity, most notably  Ag and Al at low temperatures. The predicted shifts for H atoms placed at a distance of a hundred nanometers from the walls of a Ag cavity at a temperature of 70K can be as large as 90 Hz, while their widths are of the order of a few kHz.  The obtained shift could be measurable with presently available techniques of magnetic resonance. The main experimental challenge for observing the effect is to find means of placing the atom at a well-defined  position in a metallic cavity,  for a sufficiently long time  as is necessary to measure a frequency shift of a few tens of Hz.  The experimental  investigation of this problem might shed light on open questions about the temperature dependence of dispersion forces between lossy media \cite{carsten2}.  
 
\appendix

\section{Matrix elements of the dipole operators}

In this Appendix, we compute the matrix elements of the operators $\hat{\boldsymbol{\mu}}$ and ${\hat {\bf d}}$, that are needed for the computation of the shift of ground state hyperfine levels of hydrogen. 

\subsection{Magnetic-dipole matrix elements}

We start from the magnetic-dipole moment. According to Eq. (\ref{matZee}),  the  operators ${\hat \mu}_i$ only connect the ground state hyperfine sub-levels among themselves. For brevity, we shall suppress the quantum numbers $n=1$, ${\rm L}=0$ and  ${\rm J}=1/2$ which pertain to the ground state, and thus we shall denote the state $| 1, 0,1/2;{\rm F}_a {\rm M}_a \rangle$ as $|  {\rm F}_a {\rm M}_a \rangle$.    For $L_a=0$, one has the identity 
\be
{\hat \mu}_i=g_{\rm I}\mu_{\rm n} {\hat {\rm F}}_i -(g_{\rm J} \mu_{\rm B}+g_{\rm I}\mu_{\rm n}){\hat {\rm S}}_i\;.
\ee
Using this identity, we obtain
\begin{eqnarray}
\!\!\!\!\!&&\!\!\!\!\langle{\rm F}_a {\rm M}_a | {\hat \mu}_i | {\rm F}_b {\rm M}_b \rangle =g_{\rm I}\mu_{\rm n} \langle  {\rm F}_a {\rm M}_a | {\hat {\rm F}}_i | {\rm F}_a {\rm M}_b \rangle \;\delta_{{\rm F}_a {\rm F}_b}\nonumber \\
\!\!\!\!\! &&-(g_{\rm J} \mu_{\rm B}+g_{\rm I}\mu_{\rm n}) \langle {\rm F}_a {\rm M}_a | {\hat {\rm S}}_i | {\rm F}_b {\rm M}_b \rangle \label{step1}
\end{eqnarray}
In the ground state, ${\rm F}_a$ and ${\rm F}_b$ can take the values ${\rm I} \pm 1/2$, and then two cases are possible: either ${\rm F}_a={\rm F}_b$, or ${\rm F}_a={\rm I}+1/2$ and ${\rm F}_b={\rm I}-1/2$ (the case ${\rm F}_a={\rm I}-1/2$ and ${\rm F}_a={\rm I}-1/2$ is related by hermiticity to the previous one).  
If ${\rm F}_a={\rm F}_b$,   the Wigner-Eckart theorem implies the identity
\begin{eqnarray}
&&\langle  {\rm F}_a {\rm M}_a | {\hat {\rm S}}_i | {\rm F}_a {\rm M}_b \rangle=
\tau_{{\rm F}_a} \langle  {\rm F}_a {\rm M}_a | {\hat {\rm F}}_i | {\rm F}_a {\rm M}_b \rangle \label{WigEck}
\end{eqnarray}
The number $\tau_{{\rm F}_a} $ can be computed as follows.
Upon multiplying both members by $\langle {\rm F}_a {\rm M}_c | {\hat {\rm F}}_i | {\rm F}_a {\rm M}_a \rangle$ and then summing over $i$ and $a$, one  gets
\begin{eqnarray}
\langle {\rm F}_a {\rm M}_c | {\hat {\rm {\bf F}}} \cdot  {\hat {\rm {\bf S}}}  |{\rm F}_a {\rm M}_a \rangle  
=\tau_{{\rm F}_a}\! \langle {\rm F}_a {\rm M}_a |  {\hat {\rm {\bf F}}}^2  | {\rm F}_a {\rm M}_b \rangle
\end{eqnarray}
Using the following identity, which holds for ${\rm L}_a=0$,
\be
{\hat {\rm {\bf F}}} \cdot  {\hat {\rm {\bf S}}}=\frac{1}{2} \left[{\hat {\rm {\bf F}}}^2+{\hat {\rm {\bf S}}}^2-( {\hat {\rm {\bf F}}}- {\hat {\rm {\bf S}}})^2\right]=\frac{1}{2} \left[{\hat {\rm {\bf F}}}^2+{\hat {\rm {\bf S}}}^2-{\hat {\rm {\bf I}}}^2\right]\;,
\ee
it follows that
\be
\tau_{{\rm F}_a}=\frac{\rm F(F+1)+3/4-I(I+1)}{\rm 2 \,F (F+1)}\;.\label{tau}
\ee
By combining Eq. (\ref{tau}) with Eqs. (\ref{WigEck}) and (\ref{step1}), one finds
\begin{eqnarray}
\!\!\!\!\!\!\!\!\!\!&&\!\!\!\!\langle  {\rm F}_a {\rm M}_a | {\hat \mu}_i | {\rm F}_a {\rm M}_b \rangle 
=\left[g_{\rm I}\mu_{\rm n} \frac{\rm F(F+1)-3/4+I(I+1)}{\rm 2 \,F (F+1)}  \right. \nonumber \\
\!\!\!\!\!\!\!\!\!\!&-&\!\!\! \left.g_{\rm J }\mu_{\rm B} \frac{\rm F(F+1)+3/4-I(I+1)}{\rm 2 \,F (F+1)} \right]  \!\!
 \langle {\rm F}_a {\rm M}_a | {\hat {\rm F}}_i | {\rm F}_a {\rm M}_b \rangle .
\end{eqnarray}
The matrix elements of ${\hat {\rm F}}_i$ have the well-known expressions \cite{messiah}
\begin{eqnarray}
\langle {\rm F}_a {\rm M}_a | \!\!\!\!\!\!\!\!&&\!\!\!{\hat {\rm F}}_x | {\rm F}_a {\rm M}_b \rangle=\frac{1}{2} \left(\!\sqrt{( {\rm F}_a \!-\! {\rm M}_b)( {\rm F}_a\!+\! {\rm M}_b\!+\!1)} \,\delta_{{\rm M}_a,{\rm M}_b+1} \right.\nonumber \\
&+&\left. \sqrt{( {\rm F}_a \!+\! {\rm M}_b)( {\rm F}_a\!-\! {\rm M}_b\!+\!1)} \,\delta_{{\rm M}_a,{\rm M}_b-1} \right)
\end{eqnarray}
\begin{eqnarray}
\langle {\rm F}_a {\rm M}_a | \!\!\!\!\!\!\!\!&&\!\!\!{\hat {\rm F}}_y | {\rm F}_a {\rm M}_b \rangle=\frac{1}{2{\rm i}} \left(\!\sqrt{( {\rm F}_a \!-\! {\rm M}_b)( {\rm F}_a\!+\! {\rm M}_b\!+\!1)} \,\delta_{{\rm M}_a,{\rm M}_b+1} \right.\nonumber \\
&-&\left. \sqrt{( {\rm F}_a \!+\! {\rm M}_b)( {\rm F}_a\!-\! {\rm M}_b\!+\!1)} \,\delta_{{\rm M}_a,{\rm M}_b-1} \right)
\end{eqnarray}
and
\be
\langle {\rm F}_a {\rm M}_a |  {\hat {\rm F}}_z | {\rm F}_a {\rm M}_b \rangle={\rm M}_a \;\delta_{{\rm M}_a,{\rm M}_b}\;.
\ee
Next, we consider the case ${\rm F}_a={\rm I}+1/2$ and ${\rm F}_b={\rm I}-1/2$. Then Eq.~(\ref{step1}) implies
\begin{eqnarray}
\!\!\!\!\!&&\!\!\!\!\langle  {\rm I}\!+\!1/2, {\rm M}_a | {\hat \mu}_i | {\rm I} \!-\!1/2, {\rm M}_b \rangle  =\langle  {\rm I} \!-\!1/2, {\rm M}_b | {\hat \mu}_i | {\rm I}\!+\!1/2, {\rm M}_a \rangle^* \nonumber\\
&&=-(g_{\rm J} \mu_{\rm B}\!+\!\!g_{\rm I}\mu_{\rm n}) \langle {\rm I}\!+\!1/2, {\rm M}_a |\, {\hat {\rm S}}_i\, | {\rm I} \!- \!1/2, {\rm M}_b \rangle \nonumber
\end{eqnarray}
The matrix elements of the spin operators $ {\hat {\rm S}}_i$ can be easily computed by expressing the states 
$ |{\rm F}_b {\rm M}_b \rangle$ as linear combinations of the states $ |{\rm M_I} {\rm M_S} \rangle'$ which are eigenstates of ${\hat {\rm I}}_z$ and   ${\hat {\rm S}}_z$  \cite{messiah}
\begin{eqnarray}
 |{\rm I}+1/2, {\rm M}_b \rangle &=&\!\!\left(\frac{{\rm I}+{\rm M}_b\!+\!1/2}{2 {\rm I}+1} \right)^{1/2} \!\!|{\rm M_b} \!- \!1/2, {1/2} \rangle' \nonumber\\
\!\!\!\!\! &+&\!\!\left(\frac{{\rm I}-{\rm M}_b+1/2}{2 {\rm I}+1} \right)^{1/2} \!\! | {\rm M_b} \!+\!1/2,- {1/2} \rangle' \nonumber\;,\\
\!\!\!\!\! |{\rm I}-1/2, {\rm M}_b \rangle  
 \!\!&=&-\left(\frac{{\rm I}-{\rm M}_b+1/2}{2 {\rm I}+1} \right)^{1/2}\!\! | {\rm M_b} \!-\!1/2, {1/2} \rangle' \nonumber\\
\!\!\!\!\! &+&\!\!\left(\frac{{\rm I}+{\rm M}_b \!+\!1/2}{2 {\rm I}+1} \right)^{1/2}\!\! | {\rm M_b} \!+\!1/2,- {1/2} \rangle' .\nonumber
\end{eqnarray}
By a simple algebra, one finds
\begin{eqnarray}
\!\!\!\!\!\!\!\!&&\langle {\rm I}\!+\!1/2, {\rm M}_a |\, {\hat {\rm S}}_x\, | {\rm I} \!- \!1/2, {\rm M}_b \rangle \nonumber \\
\!\!\!\!\!\!\!\!&&=\frac{\sqrt{ ({\rm I}+{\rm M}_b+3/2)({\rm I}+{\rm M}_b+1/2) }}{2(2 {\rm I}+1)} \delta_{{\rm M}_a,{\rm M}_b+1}\nonumber \\
&&-\frac{ \sqrt{({\rm I}-{\rm M}_b+3/2)({\rm I}-{\rm M}_b+1/2) }}{2(2 {\rm I}+1)} \delta_{{\rm M}_a,{\rm M}_b-1} \nonumber
\end{eqnarray}
\begin{eqnarray}
\!\!\!\!\!\!\!\!&&\langle {\rm I}\!+\!1/2, {\rm M}_a |\, {\hat {\rm S}}_y\, | {\rm I} \!- \!1/2, {\rm M}_b \rangle \nonumber \\
\!\!\!\!\!\!\!\!&&=\frac{1}{\rm i} \left\{\frac{\sqrt{ ({\rm I}+{\rm M}_b+3/2)({\rm I}+{\rm M}_b+1/2) }}{2(2 {\rm I}+1)} \delta_{{\rm M}_a,{\rm M}_b+1} \right.\nonumber \\
&&\left.+\frac{ \sqrt{({\rm I}-{\rm M}_b+3/2)({\rm I}-{\rm M}_b+1/2) }}{2(2 {\rm I}+1)} \delta_{{\rm M}_a,{\rm M}_b-1}\right\}\nonumber
\end{eqnarray}
\begin{eqnarray}
\langle {\rm I}\!+\!1/2, {\rm M}_a |\, {\hat {\rm S}}_z\, | {\rm I} \!- \!1/2, {\rm M}_b \rangle 
\!\!=\!\! -\frac{\sqrt{ ({\rm I}\!+\! 1/2)^2-{\rm M}^2_b  }}{(2 {\rm I}+1)} \delta_{{\rm M}_a,{\rm M}_b} \nonumber
\end{eqnarray}

\subsection{Electric-dipole matrix elements}

In this Section we work out the matrix elements of the electric-dipole moment operators ${\hat d}_i$ that are needed for the computation of the Stark effect of ground-state hyperfine levels. The operators ${\hat d}_x$, ${\hat d}_y$ and ${\hat d}_z$ obey the selection rule $\Delta {\rm L}= \pm 1$, and thus they connect the ground-state hyperfine sub-levels, which have ${\rm L}=0$, with  excited states having ${\rm L}=1$
\be
d_{i}^{ab}=\langle 1, 0 ,1/2;{\rm F}_a{\rm M}_a | {\hat d}_i | n_b, 1,{\rm J}_b; {\rm F}_b {\rm M}_b \rangle \;.
\ee
To compute these matrix elements, it is convenient to first  factorize $d_{i}^{ab}$ into radial and an angular parts,
\be
d_{i}^{ab} = - e \,a_0 \, R_{n_b 1}^{10}\, \left\langle 1, 0 ,1/2;{\rm F}_a{\rm M}_a \left| \frac{{\hat x}_i}{r} \right| n_b, 1,{\rm J}_b; {\rm F}_b {\rm M}_b \right\rangle
\ee
where $a_0$ is the Bohr radius, $r$ is the radial coordinate, and $R_{n_b {\rm L}_b}^{n_a {\rm L}_a}$ denote the radial integrals
\be
R_{n_b {\rm L}_b}^{n_a {\rm L}_a}= \frac{1}{a_0}\int_0^{\infty} dr\,r^3\,R^*_{n_b {\rm L}_b}(r) R_{n_a{\rm L}_a}(r)\;.
\ee
Here $ R_{n_a{\rm L}_a}(r)$ and  $R_{n_b{\rm L}_b}(r)$ are the radial parts of the atomic wave function. For hydrogen the integrals $R_{n_b {\rm L}_b}^{n_a {\rm L}_a}$ can be expressed in terms of confluent  hypergeometric functions \cite{bethe}.   For  discrete energy states they have the simple expression
\be
\left|R^{10}_{n1}\right|=16 \,\sqrt{\frac{n^7\,(n-1)^{2n-5}}{(n+1)^{2n+5}}}\;.
\ee
The radial integrals are more involved for continuum states.   States of energy $E>0$ can be   labelled by the positive real number ${\tilde k}=k a_0$, where $k=\sqrt{2 m E}/\hbar$. With this parametrization,  the radial wave functions $R_{{\tilde k} l}(r)$ of continuum states are normalized  \footnote{Our normalization coincides with the $k$-scale of \cite{bethe}} such that
\be
\langle R_{{\tilde k}'\, l} | R_{{\tilde k}\, l}\rangle=\int_0^{\infty} dr\,r^2\,R^*_{{\tilde k}' l}(r) R_{{\tilde k}\,l}(r)= \delta({\tilde k}'-{\tilde k})\;.
\ee
Then one finds
\begin{eqnarray}
&&\left|R^{10}_{{\tilde k} 1} \right|= \left|\frac{1}{a_0}\int_0^{\infty} dr\,r^3\,R^*_{{\tilde k} 1}(r) R_{1\,0}(r)\right| \nonumber \\
&&=\frac{4}{3}\; \frac{1}{{\tilde k}^4} \sqrt{\frac{{\tilde k}+1/{\tilde k}}{1-\exp(-2 \pi/{\tilde k})}} \; \left|I({\tilde k}) \right|\;,
\end{eqnarray}
where $I({\tilde k})$ is the integral
\be
I({\tilde k})=\int_0^{\infty} dx \;x^4 \;e^{-({\rm i}  +1/{\tilde k}) x }\; \Phi(2+{\rm i}/{\tilde k}; 4; 2 \,{\rm i}\, x)\;. \label{intI0}
\ee
Here, $\Phi(a;c;x)$ is the {\it confluent hypergeometric function}. The integral $I({\tilde k})$ can be expressed in terms of 
the hypergeometric function ${\,}_2F_1(a,b;c;z)$. The lenghty formula is omitted for brevity.

Now, we consider the matrix elements of the angular operators  ${\hat d}_i/r$. They can be  expressed as
\begin{eqnarray}
\frac{{\hat x}}{r}&=&  \frac{{\hat C}^{1}_{-1}-{\hat C}^{1}_{1}}{\sqrt 2}\;,\nonumber\\
\frac{{\hat y}}{r}&=& {\rm i} \,\frac{ {\hat C}^{1}_{-1}+{\hat C}^{1}_{1}}{\sqrt 2}\;,\nonumber\\
\frac{{\hat z}}{r}&=& \,{\hat C}^{1}_{0}\;,
\end{eqnarray}
where   ${\hat C}^{l}_{q}$ denote the multiplication operators by the normalized spherical harmonics 
\be
{C}^{l}_{q}:=\sqrt{\frac{4 \pi}{2 l+1}}\; Y^{l}_q
\ee

To compute the matrix elements of the operators ${\hat C}^{l}_{q}$, it is useful  to introduce two more sets of states. Let ${\hat {\bf G}}={\hat {\bf S}}+{\hat {\bf I}}$ be the sum of the electron and nuclear spin operators. We let  $|n, {\rm L},{\rm G};{\rm F} {\rm M} \rangle'$ the states which are eigenstates of ${\hat {\bf G}}^2$. Moreover, we consider the states  $|n; {\rm L},{\rm M}_{\rm L};{\rm G}{\rm M}_{\rm G} \rangle''$ that are eigenstates of ${\hat {\rm L}}_z$, ${\hat {\bf G}}^2$ and  ${\hat {\rm G}}_z$. Since for ${\rm L}=0$ it holds  ${\rm G}={\rm F}$ and ${\rm M}_{\rm G}={\rm M_F}$ it follows that
\begin{eqnarray}
\!\!\!\!\!\!\!\!&&\!\!\!\!\!\!\!\!{\hat C}^{1}_{q}| 1, 0 ,1/2;{\rm F}_a{\rm M}_a\rangle={\hat C}^{1}_{q} | 1; 0 ,0;{\rm F}_a{\rm M}_a\rangle'' \nonumber \\
\!\!\!\!\!\!\!\!&=& \!\!\frac{1}{\sqrt 3} \;| 1; 1 ,q;{\rm F}_a{\rm M}_a\rangle'' \nonumber \\
\!\!\!\!\!\!\!\!&=&\!\!\frac{1}{\sqrt 3}  \sum_{{\rm F}_c,{\rm M}_c} |n, 1,{\rm F}_a;{\rm F}_c {\rm M}_c \rangle' \langle {\rm F}_c {\rm M}_c| 1,q; {\rm F}_a {\rm M}_a \rangle, \label{elmat1}
\end{eqnarray}
where $\langle {\rm F}_c {\rm M}_c| 1,q; {\rm F}_a {\rm M}_a \rangle$ is a Clebsch-Gordan coefficient.
At this point, using the ``6j'' symbols \cite{messiah} we express the state $ | n_b, 1,{\rm J}_b; {\rm F}_b {\rm M}_b \rangle$ as a combination of the states
$|n_b, 1,{\rm G};{\rm F}_b {\rm M_b} \rangle'$
\begin{eqnarray}
&& | n_b, 1,{\rm J}_b; {\rm F}_b {\rm M}_b\rangle \nonumber\\
&&=\sum_{\rm G}  | n_b, 1,{\rm G}; {\rm F}_b {\rm M}_b\rangle' \, \sqrt{(2 {\rm J}_b+1)(2 {\rm G}+1)} \nonumber \\
&&\times(-1)^{3/2+{\rm I}+{\rm F}_b} \left\{ \begin{array}{c}
\!{\rm 1\;\;1/2\;\;{\rm J}_b} \\
\!{\rm I}\;\; \;\;{\rm F}_b\;\;\;{\rm G} 
\!\end{array}\right\}\;.\label{elmat2}
\end{eqnarray}
By combining Eq. (\ref{elmat1}) and (\ref{elmat2}) we obtain
\begin{eqnarray}
\!\!\!\!\!\!\!\!&&\langle 1, 0 ,1/2;{\rm F}_a{\rm M}_a | {\hat C}^{1}_q | n_b, 1,{\rm J}_b; {\rm F}_b {\rm M}_b \rangle \nonumber \\&=&
\! \frac{1}{\sqrt{3}}\sqrt{(2 {\rm J}_b+1)(2 {\rm F}_a+1)} (-1)^{3/2+{\rm I}+{\rm F}_b}
\nonumber\\
&&\times \left\{ \begin{array}{c}
\!{\rm 1\;\;\;1/2\;\;\;{\rm J}_b} \\
\!{\rm I}\;\; \;\;{\rm F}_b\; \;\;{\rm F}_a 
\!\end{array}\right\}  \langle 1,q;{\rm F}_a {\rm M}_a |{\rm F}_b {\rm M}_b \rangle\;.
\end{eqnarray}


\section{ Green function of a planar cavity}

In this Appendix we provide the explicit formulae for the Green functions of a cavity, that are needed for the computations described in the present work.

At  points $\bf r$ and ${\bf r}'$ in the gap between two parallel dielectric slabs at distance $a$  in vacuum,  the electric Green function ${\cal E}_{ij}({\bf r},{\bf r'},\omega)$  can be decomposed as
\be
{\cal E}^{(\rm cav)}_{ij}({\bf r},{\bf r'},\omega)={\cal E}^{(0)}_{ij}({\bf r},{\bf r'},\omega)+{\cal E}^{(\rm sc)}_{ij}({\bf r},{\bf r'},\omega)\;, \label{Greensplit}
\ee
where ${\cal E}^{(0)}_{ij}({\bf r},{\bf r'},\omega)$ is the free-space Green function, and ${\cal E}^{(\rm sc)}_{ij}({\bf r},{\bf r'},\omega)$ is a scattering contribution. The magnetic Green function has an analogous decomposition
\be
{\cal H}^{(\rm cav)}_{ij}({\bf r},{\bf r'},\omega)={\cal H}^{(0)}_{ij}({\bf r},{\bf r'},\omega)+{\cal H}^{(\rm sc)}_{ij}({\bf r},{\bf r'},\omega)\;. \label{GreensplitH}
\ee
The free-space Green function has the expression
$$
{\bf \cal E}^{(0)}({\bf r},{\bf r}',\omega)={\bf \cal H}^{(0)}({\bf r},{\bf r}',\omega)=\left[(3 {\hat {\bf R}} \otimes {\hat {\bf R}} - {\bf 1}) \left( \frac{1}{R^3}- \frac{i\,\omega}{c R^2}\right) \right.
$$ 
\be
+\left. ({\bf 1}- {\hat {\bf R}} \otimes {\hat {\bf R}}) \frac{\omega^2}{c^2 R} - \frac{4 \pi}{3} \delta({\bf R}){\bf 1}\right]e^{i \omega R/c}\;,\label{Green0}
\ee
where ${\bf R}= {\bf r}-{\bf r}'$. Note that the imaginary part of the free-space Green function is non-singular for ${\bf r} \rightarrow {\bf r}'$,
\be
\lim_{{\bf r} \rightarrow {\bf r}'}{\rm Im}[{\bf \cal E}^{(0)}({\bf r},{\bf r}',\omega)]=\lim_{{\bf r} \rightarrow {\bf r}'}{\rm Im}[{\bf \cal H}^{(0)}({\bf r},{\bf r}',\omega)]=\frac{2 \omega^3}{3 c^3}{\bf 1}\;.
\ee
In the limit   ${\bf r} \rightarrow {\bf r}'$, the scattering part of the Green tensor ${\cal E}^{(\rm sc)}_{ij}({\bf r},{\bf r}',\omega)$ attains a finite limit, and its non vanishing components are
$$
{\cal E}^{(\rm sc)}_{y y}({\bf r},{\bf r},\omega)={\cal E}^{(\rm sc)}_{z z }({\bf r},{\bf r},\omega)=4 \pi i \int \frac{d^2 {\bf k}_{\perp}}{(2 \pi)^2} k_x 
$$
$$
\times \left[\left(\frac{R_{\rm p}^{(1)}  R_{\rm p}^{(2)}}{{\cal A}_{\rm p}}+ \frac{\omega^2}{c^2 k_x^2}\frac{R_{\rm s}^{(1)}  R_{\rm s}^{(2)}}{{\cal A}_{\rm s}}  \right) e^{2 i k_x a} \right. + \frac{1}{2}\left( \frac{\omega^2}{c^2 k_x^2} \frac{R_{\rm s}^{(1)} }{{\cal A}_{\rm s}} \right.
$$
\be
\left.\left. - \frac{R_{\rm p}^{(1)} }{{\cal A}_{\rm p}}\right)e^{2 i k_x x}+ 
 \frac{1}{2}\left( \frac{\omega^2}{c^2 k_x^2} \frac{R_{\rm s}^{(2)} }{{\cal A}_{\rm s}} - \frac{R_{\rm p}^{(2)} }{{\cal A}_{\rm p}}\right)e^{2 i k_x (a-x)}\right]\;, \label{Exx}
\ee
and
$$
{\cal E}^{(\rm sc)}_{x x}({\bf r},{\bf r},\omega)= 4 \pi i \int \frac{d^2 {\bf k}_{\perp}}{(2 \pi)^2} \frac{k_{\perp}^2}{k_x} \left( \frac{R_{\rm p}^{(1)}  R_{\rm p}^{(2)}}{{\cal A}_{\rm p}} e^{2 i k_x a} \right.
$$
\be
\left. + \frac{R_{\rm p}^{(1)} }{2 {\cal A}_{\rm p}}\, e^{2 i k_x x} +  \frac{R_{\rm p}^{(2)} }{2 {\cal A}_{\rm p}} \,e^{2 i k_x (a-x)} \right)\;,\label{Ezz}
\ee
where ${\bf k}_{\perp}$ is the in-plane wave-vector, $k_x=\sqrt{\omega^2/c^2-k_{\perp}^2}$,  the indices $\rm s$ and $\rm p$ denote TE and TM polarizations, respectively, $R_{\alpha}^{(k)}$  is the reflection coefficient of the $k$-th mirror for polarization $\alpha={\rm s}, {\rm p}$ and  ${\cal A}_{\alpha}=1-R_{\alpha}^{(1)} R_{\alpha}^{(2)} e^{2 i k_x a}$. The corresponding formulae for the magnetic Green tensor  ${\cal H}^{(\rm sc)}_{ij}({\bf r},{\bf r},\omega)$ can be obtained from those of the electric Green tensor, by interchanging the reflection coefficients  $R_{\rm s}^{(k)} \leftrightarrow R_{\rm p}^{(k)}$ into Eqs. (\ref{Exx}) and (\ref{Ezz}).

As shown in Fig. \ref{Fig3}, our mirrors consist of a metallic layer of thickness $w$, deposited on a thick Si substrate.  The reflection coefficient   $R_{\alpha}$ of a layered mirror with such a structure  has the expression
\be  
R_{\alpha}( \omega,k_{\perp};w)=\frac{r_{\alpha}^{(0\,{\rm met})}+e^{2{\rm i }\,w\, k_x^{({\rm met})}}\,r_{\alpha}^{({\rm met\, Si})}}{1+e^{2 {\rm i}\,w\, k_x^{({\rm met})}}\,r_{\alpha}^{(0\,{\rm met})}\,r_{\alpha}^{({\rm met\, Si})}}\;.\label{refl1}
\ee   
Here $r^{(ab)}_{\alpha}$ are the Fresnel reflection coefficients for a planar dielectric (we set $\mu=1$ for all materials) interface separating medium a from medium b,
\be
r^{(ab)}_{\rm TE}=\frac{  \,k_x^{(a)}-  \,k_x^{(b)}}{ k_x^{(a)}+ \,k_x^{(b)}}\;,\label{freTE}
\ee
\be
r^{(ab)}_{\rm TM}=\frac{\epsilon_{b} (\omega) \,k_x^{(a)}-\epsilon_{a}(\omega) \,k_x^{(b)}}{\epsilon_{b}(\omega) \,k_x^{(a)}+\epsilon_{a}(\omega) \,k_x^{(b)}}\;,\label{freTM}
\ee
where
$ k_x^{(a)}=\sqrt{\epsilon_a(\omega)  \omega / c^2-k_{\perp}^2}\;$,   and $\epsilon_{a}(\omega)$   is the complex frequency-dependent permittivity of medium $a$.

\end{document}